\pdfoutput=1
\documentclass[12pt]{article}
\usepackage{hyperref}   
\usepackage{amsmath}
\usepackage{slashed}
\usepackage{amssymb}
\usepackage{euscript}
\usepackage{bbm}
\usepackage{bm}
\usepackage[usenames]{color}
\usepackage[T1]{fontenc}
\usepackage[utf8]{inputenc}

\usepackage{graphicx}

\newcommand{\Peu}{\EuScript{P}}

\newcommand{\Pbbm}{\mathbbm{P}}
\newcommand{\Rbbm}{\mathbbm{R}}

\newcommand{\Bbbm}{\mathbbm{B}}
\newcommand{\Pop}{\overleftarrow{\Pbbm}}

\newcommand{\Bop}{\overleftarrow{\Bbbm}}
\newcommand{\Rop}{\overleftarrow{\Rbbm}}
\newcommand{\ba}{{\bf{a}}}
\newcommand{\bk}{{\bf{k}}}

\begin{document}

\begin{center}
\vspace{-20mm}
\begin{flushright} {\bf IFJPAN-IV-2009-9}\\
\end{flushright}

\vspace{5mm}
{\Large Exclusive Monte Carlo modelling\\ of NLO DGLAP evolution}
\footnote{
  This work is partly supported by the EU 
  Framework Programme grants MRTN-CT-2006-035505 and
  MTKD-CT-2004-014319
  and by the Polish Ministry of Science and Higher Education grants
  No. N N202 128937 and 153/6.PR UE/2007/7.}%

\vspace{4mm}
{\large S. Jadach and M. Skrzypek and A. Kusina and M. Sławińska}\\
\vspace{4mm}
%
{\em H. Niewodniczański Institute of Nuclear Physics, \\Polish
  Academy of Sciences,\\
  ul.\ Radzikowskiego 152, 31-342 Cracow, Poland,} 
\end{center}


\vspace{2mm}
\begin{abstract}
{\em Abstract:}
The next-to-leading order (NLO) evolution of the parton distribution functions
(PDFs) in QCD is a common tool in the lepton-hadron and hadron-hadron
collider data analysis. The standard NLO DGLAP evolution is formulated for
inclusive (integrated) PDFs and done using inclusive NLO kernels. We report
here on the ongoing project, called KRKMC, in which NLO DGLAP evolution is
performed for the exclusive multiparton (fully unintegrated) distributions
(ePDFs) with the help of the exclusive kernels. These kernels are calculated
within the two-parton phase space for the non-singlet evolution, using Curci-Furmanski-Petronzio factorization scheme. 
The multiparton distribution, with multiple use of the
exclusive NLO kernels, is implemented in the Monte Carlo program simulating
multi-gluon emission from single quark emitter. High statistics tests
($\sim 10^{10}$ events) show that the new scheme works perfectly well in
practice and, at the inclusive (integrated) level,
is equivalent with the traditional inclusive NLO DGLAP evolution. 
Once completed, this new technique is aimed as a
building block for the new more precise
NLO parton shower Monte Carlo, for W/Z production at LHC
and for ep scattering, as well as a starting point for other perturbative QCD
based Monte Carlo projects.

\vspace{3mm}
\centerline{\em Presented by S.~Jadach at RADCOR 2009 Int. Symposium,}
\centerline{\em Oct. 25-30, 2009, Ascona, Switzerland}
\end{abstract}

\vspace{5mm}
\begin{flushleft}
\bf IFJPAN-IV-2009-9\\
\end{flushleft}


\vfill\newpage

We report on the ongoing effort on
the exclusive Monte Carlo (MC) modeling of DGLAP\cite{DGLAP}
evolution at the NLO level using classic work of
Curci Furmanski and Petronzio (CFP)~\cite{Curci:1980uw}
as a guide and reference.

The so-called
{\em factorization theorems}~\cite{Ellis:1978sf,Collins:1984kg,Bodwin:1984hc}
in Quantum Chromodynamics (QCD)
are stating that in the high energy scattering process of hadrons,
with an experimentally identifiable single large scale
(effective mass, transverse momentum etc.) one may reorganize the infinite
order perturbative expansion in terms of Feynman diagrams, such that
all collinear (mass) singularities are encapsulated into certain well
defined objects, called parton distribution functions (PDFs)
or parton fragmentation functions (PFFs or jets),
while the remaining part, free of such singularities,
forms the so-called hard process part (coeff. function).
The soft singularities due to zero mass gluon emissions are shown not
to disturb or invalidate this picture~\cite{Collins:1984kg}, 
if they are averaged/integrated over the phase space and properly
combined with the virtual contributions.
In the physical gauge the PDF/PFF part consists of
a well defined Feynman diagrams with the ladder topology.
In the early stage of formulating practical QCD perturbative
methodology it was found that the most economical way of dealing with
the PDF/PFF parts of the process was to define them
as {\em inclusive} as possible,
integrating over transverse momenta and summing up over
all partons emitted from the ladder, keeping control only
on the total energy (light-cone variable) of the parton
entering the hard process, and its type.
Such inclusive (collinear) PDF is widely used until today
in most of practical QCD calculations, especially for the initial hadrons.

The only exception is the so-called parton shower Monte Carlo (PSMC),
where one gains access to all momenta and other quantum numbers in
PDF/PFF (ladder) part of the hadronic scattering, for every incoming
hadron or outgoing jet.
Originally the main role of PSMCs was to describe hadronization of the partons,
but they have gradually absorbed the leading order
perturbative QCD (pQCD)
description~\cite{Sjostrand:1985xi,Marchesini:1988cf}
of the ladders (PDFs, PFFs).
With the growing sophistication of the high energy (HE) experimental detectors
PSMC became indispensable for understanding data in any modern experiment.
However, although pQCD calculations using inclusive PDFs have evolved
enormously in their sophistication
(evolution of PDFs at NLL, NNLL level,
corrections to hard process at NLO, NNLO, new ingenious methods of
calculating tree-level multiple parton distributions and more)
the PSMCs have stayed, from the pQCD point of view,
where they were 25 years ago,
that is at the (improved) LO/LL level, until today.
This lack of the progress is not fully understood,
but most likely the main reason was that computers fast enough
were not available and due to the difficulties
in reformulating QCD factorization theorems into a form suitable
for stochastic simulation (MC) methods.

We are reporting on the first serious attempt to
upgrade parton shower MC for a single incoming quark (non-singlet PDF)
to the level of the {\em complete} NLO%
\footnote{See also
  refs.~\cite{Tanaka:2003gk,Collins:2000gd,Collins:2007ph}
  for similar effort in this direction.}.
This will be done, as in early days of pQCD, in the physical gauge,
including first order real and virtual corrections
to the basic ladder describing LO level showering of one incoming quark.
Our Monte Carlo implementation of NLO DGLAP evolution is:
(1)
based firmly on Feynman diagrams and standard LIPS,
(2)
based rigorously on the collinear factorization
(eg. EGMPR~\cite{Ellis:1978sf}),
(3)
implementing {\em exactly} NLO $\overline{MS}$ DGLAP evolution
at the inclusive level,
(5)
defining fully unintegrated exclusive ePDFs
(the integrand of inclusive PDFs),
(6)
performing NLO evolution by the MC itself
(no use of backward evolution\cite{Sjostrand:1985xi}).

\begin{figure}[!ht]
\centering
{\includegraphics[width=70mm]{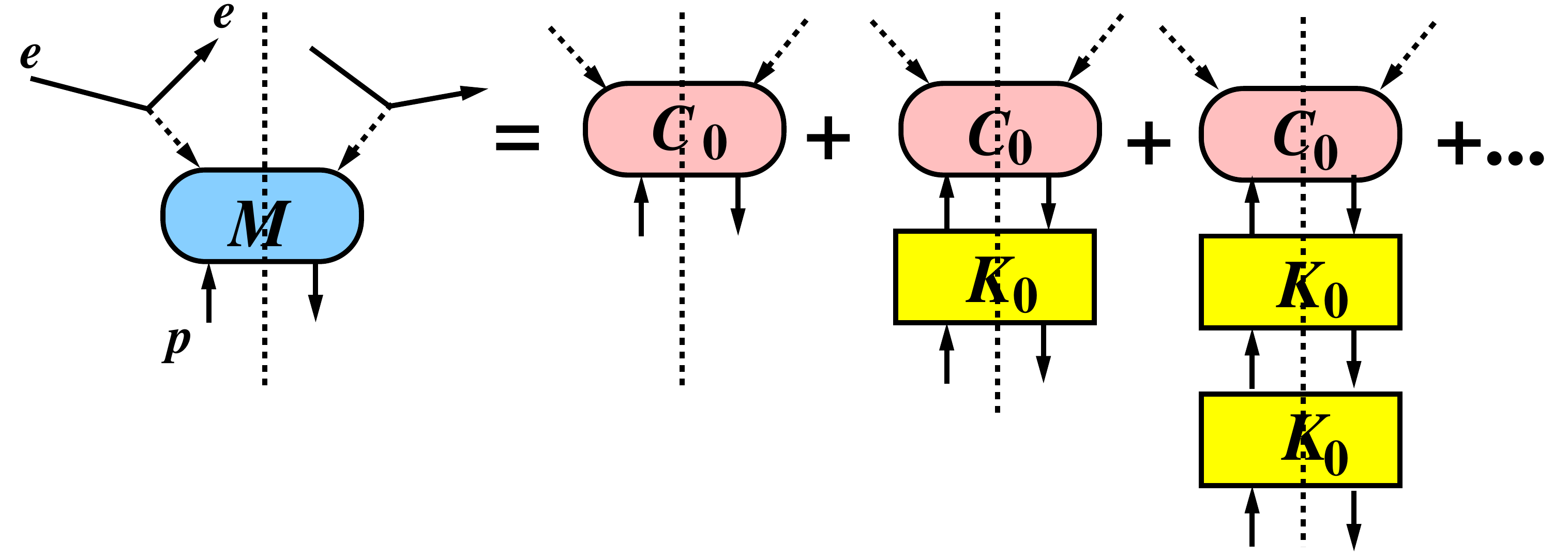}}
\caption{
   Raw factorization theorem in the physical gauge.
}
\label{fig:one}
\end{figure}

Factorization scheme of EGMPR~\cite{Ellis:1978sf}, see Fig.~\ref{fig:one},
was customized to $\overline{MS}$ by CFP~\cite{Curci:1980uw}:
\begin{equation}
\begin{split}
F&=C_0\cdot  \frac{1}{1-K_0}
=C\left(\alpha,\frac{Q^2}{\mu^2} \right)
\otimes \Gamma\left(\alpha, \frac{1}{\epsilon}\right),
\\
\Gamma&\left(\alpha, \frac{1}{\epsilon}\right)
 \equiv \left(\frac{1}{1- K}\right)_\otimes
 = 1+K+K\otimes K +K\otimes K\otimes K+...,
\\
K&=\Pbbm K_0\cdot\frac{1}{1-(1-\Pbbm)\cdot K_0},
\quad
C=C_0\cdot \frac{1}{1-(1-\Pbbm)\cdot K_0},
\end{split}
\end{equation}
\small
where the ladder part $\Gamma$ corresponds to MC parton shower
and $C$ is the hard process part
\footnote{
  Multiplication symbol $\cdot$ means
  full phase space integration $d^n k$
  while convolution $\otimes$ only the integration
  over the 1-dim. lightcone variable.}.
NLO kernels were extracted in ref.~\cite{Curci:1980uw}
from the coefficient of a single pole $\frac{1}{\epsilon}$ in $\Gamma$.
Projection operator of CFP,
$
 \Pbbm = P_{spin}\; P_{kin}\; PP,
$
consists of the kinematic (on-shell) projection operator $P_{kin}$,
spin projection operator $P_{spin}$
and $PP$ extracting pole part $\frac{1}{\epsilon^k}, k>0$.

In our MC solutions we use
the standard interpretation~\cite{Frixione:2002ik}
of the collinear $\epsilon$-poles:
\begin{equation}
\frac{1}{\varepsilon}=
\int_0^{\mu_F} \frac{d k^T}{k^T}\;
\left(\frac{k^T}{\mu_F}\right)^\epsilon.
\end{equation}
However, the ladder part in CFP/EGMPR scheme
features enormous cancellations,
as can be seen already at the LO level%
\footnote{Omitting for simplicity $1/\epsilon^n$ poles due to running of 
   the coupling constant from the consideration.}:
\begin{equation}
\Gamma \simeq \frac{1}{1-\Big(1-e^{-\frac{1}{\varepsilon}}\Big)}
=1+\Big(1-e^{-\frac{1}{\varepsilon}}\Big) +\Big(1-e^{-\frac{1}{\varepsilon}}\Big)^2+... ,
\end{equation}
while from RGE and explicit LO calculation we obtain readily
$
\Gamma=e^{+\frac{1}{\varepsilon}}=1
      +\frac{1}{\varepsilon}+\frac{1}{2!}\frac{1}{\varepsilon^2}+...
$
In the MC we need this exponent manifestly,
if possible directly from the Feynman diagrams!

The above exponential nature of the QCD evolution of PDFs is manifest in
the following master formula
\begin{equation}
\label{eq:master}
\begin{split}
&F=C_0\cdot\frac{1}{1-K_0}
=C_0 \cdot \Rop_\mu[K_0] \cdot
\exp_{TO} \left( \Pop' \left\{^s K_0 \cdot \Rop_s[K_0] \right\}
\right)_{\mu},
\\&
\Rop_\mu(K_0) = \Bop_\mu \left[ \frac{1}{1-K_0} \right]
\equiv 1+\Bop_\mu[K_0]+\Bop_\mu[K_0 \cdot K_0]
  +\Bop_\mu[K_0 \cdot K_0 \cdot K_0]+\dots
\end{split}
\end{equation}
which is serving as a generating functional
of the exclusive parton distributions implemented in the MC.
Here,
$\exp_{TO}$ is {\em the time ordered exponential}
in the time evolution variable $t=\ln \mu$,
where $\mu$ is factorization scale variable.
Operator $\Bop$ is defined recursively%
\footnote{
  Similarly as $\beta$-functions in Yennie-Frautschi-Suura\cite{yfs:1961}
  subtraction scheme. See also \cite{Collins:1998rz}.}:
\begin{equation}
\begin{split}
&\Bop_\mu [ K_0 ]  = K_0 - \Pbbm'_\mu\{^s K_0 \},
\\&
\Bop_\mu [ K_0 \cdot K_0 ]\!  =\!
K_0 \cdot K_0 \!
 -\Pbbm'_\mu\{^{s_2} K_0 \}\! \cdot \Pbbm'_{s_2}\{^{s_1} K_0 \}\!
 -\Pbbm'_\mu\{^{s_2} K_0 \!   \cdot \Bop_{s_2}[K_0] \}\!
 -\Bop_\mu[ K_0 ]\!           \cdot \Pbbm'_{\mu}\{ K_0 \},
\\&
\Bop_\mu [ K_0 \cdot K_0 \cdot K_0 ]  =
K_0 \cdot K_0 \cdot K_0 
  -\Pbbm'_\mu\{^{s_3} K_0 \}
   \cdot \Pbbm'_{s_3}\{^{s_2} K_0 \}
   \cdot \Pbbm'_{s_2}\{^{s_1} K_0 \}
  -\dots
\end{split}
\end{equation}
More terms in the recursion can be obtained by expanding LHS and RHS of
eq.~(\ref{eq:master}) in powers of $K_0$ and using definition of
$\exp_{TO}$ (see below).
The key point is the definition of modified projection operator $\Pop'$:
(a)
it does spin projection as $\Pbbm$ of CFP,
(b)
it sets its incoming momentum on-shell in the part
of the diagram {\em towards} the hard process,
(c)
it acts on the integrand
of the Lorentz invariant phase space (LIPS), before integration,
(d)
it sets upper limit $\mu$ on the phase space for all
{\em its own} real (cut) partons, eg. $\mu>s(k_1,..,k_n)=\max(k^T_i)$,
(e)
our preferred choice is {\em rapidity ordering} choice;
$s(k_1,..,k_n)=a(k_1,..,k_n)=\max(k^T_i/\alpha_i)$, $\alpha_i=k^+_i/E$,
(f)
$\Pop'_\mu(A)$ acts on $A$ which is {\em at most}
single-log (collinear) divergent
and extracts this singularity from the LIPS integrand%
\footnote{%
    For instance by rescaling all $k^T_i\to \lambda k^T_i$ 
    and taking coefficient in front of $1/\lambda$ term.},
(g)
$\Pop'_\mu(K_0)$ is legal, as $K_0$ is single-log divergent,
(h)
nesting like $\Pop'[K_0\cdot(1- \Pop'(K_0))]$ is allowed, as long
as its argument is no more than single-log divergent,
(e) $\Pop'$ does not include PP operation.
Finally, the time ordered exponential reads:
\begin{equation}
\exp_{TO}\left( \Pbbm'_\mu\{ A \} \right)_{\mu}
\!=\!
1 +\Pbbm'_\mu\{ A \}
  +\Pbbm'_\mu\{^{s_2} A \}\cdot\Pbbm'_{s_2}\{^{s_1} A \}
  +\Pbbm'_\mu\{^{s_3} A \}\cdot
   \Pbbm'_{s_3}\{^{s_2} A \}\cdot\Pbbm'_{s_2}\{^{s_1} A \}
  +\dots
\end{equation}
where notation {$\{^{s} A \}$} means that
$s=a(a_1,...,a_n)=\max(a_1,...,a_n)$.
For instance for $n=3$ the entire integrand multiplied by
$\theta_{\mu>s_3>s_2>s_1}$.
Variable $\mu$ is constant, while $s_i$ depend
on the 4-momenta integration variables.

Master formula of eq.~(\ref{eq:master}) is very important for
us, as it serves as a generating functional of the exclusive
distributions beyond LO implemented in the MC.
In CFP scheme the time ordered exponential
is present for the inclusive PDFs, and results
from the renormalization group equations,
not directly from Feynman diagrams.
Deriving eq.~(\ref{eq:master}) directly from
diagrams at any perturbative order remains an open important problem%
\footnote{%
   In particular the effect
   of the running coupling costant deserves careful discussion.}.
For the time being we check order by order that:
(i) at the inclusive level we maintain full compability with
the CFP scheme,
(ii) we reproduce fixed order matrix squared element times LIPS
results wherever possible and/or necessary.

From now on,
in the factorization formula (\ref{eq:master}),
we focus on the exclusive PDF (ePDF) $\mathcal{D}(\mu,k_1,...,k_n)$,
which is the integrand of the inclusive PDF:
\begin{equation}
D(\mu)=
\exp_{TO}\left(\Pop'\left\{^s K_0 \cdot\Rop_s[K_0]\right\}\right)(\mu)
=\exp_{TO}(K),
\end{equation}
The $x$-dependent PDFs (inclusive) are obtained
by means of inserting $\delta(x-x(k_1,...,k_n))$ in the integrand
\footnote{Such an insertion we shall often mark as $(...)_x$.
  In the MC it means histogramming of $x$.
  It also helps eliminating $x=0$ from the consideration.},
$D(\mu)\to D(\mu,x)=D(\mu)_x$.

The standard inclusive PDF,
$D(\mu,x)=\int d{\rm Lips}\; \mathcal{D}(\mu, k_1,...,k_n, x)$,
is obtained form ePDF by the phase space integration.
It obeys by construction the ordinary evolution equation
\begin{equation}
  \partial_\mu D(\mu,x) = \Peu\otimes D(\mu,x)
\end{equation}
with the inclusive DGLAP kernel
\begin{equation}
\begin{split}
\Peu(x) &
= \frac{\partial}{\partial\ln(\mu)} (K_\mu)_x\;
= \int\!\! d{\rm Lips}\; \delta\bigg(x-\frac{\sum k_i^+}{E_0}\bigg)\;
  \delta\left(1-\frac{s}{\mu}\right)
  \frac{d}{d{\rm Lips}}
  \Pop'_\mu \left\{^s K_0 \cdot \Rop_s[K_0] \right\}.
\end{split}
\end{equation}
The LO and NLO truncations of the evolution kernel $K_\mu$ are:
$ K^{LO}_\mu =\Pop'_\mu \left\{^s K_0 \right\}$,
taken at ${\cal O}(\alpha^1)$
and $ K^{NLO}_\mu
  =\Pop'_\mu \left\{^s \big(K_0 +K_0\cdot (1-\Pop')\cdot K_0 \big)\right\}$,
truncated at ${\cal O}(\alpha^2)$.
The 2PI kernel $K_0$ of CFP scheme (non-singlet bremsstrahlung) at LO+NLO is:
\begin{equation}
\label{eq:K0NLO}
K_0=
2\Re\bigg(
\raisebox{-11pt}{\includegraphics[height=11mm]{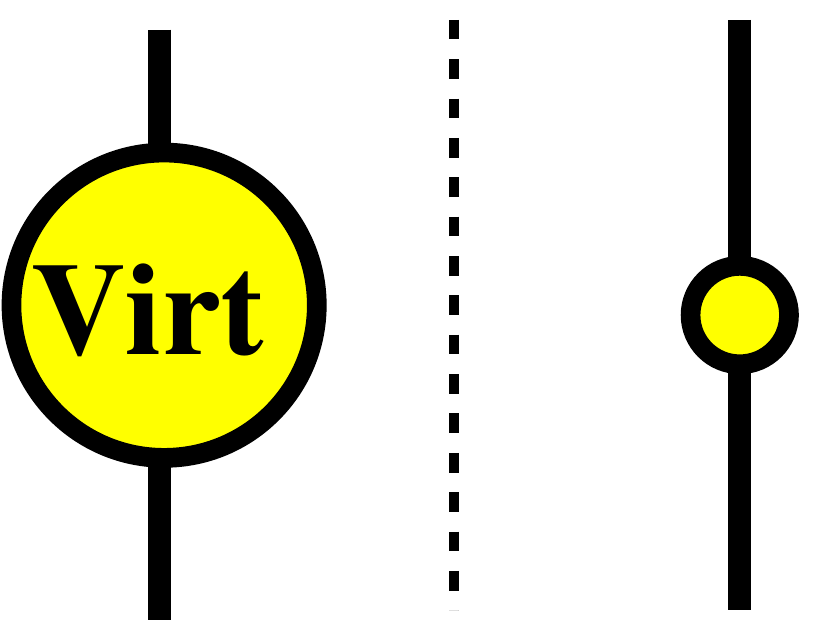}} \bigg)
\hbox{\LARGE +}\;\;
\raisebox{-11pt}{\includegraphics[height=11mm]{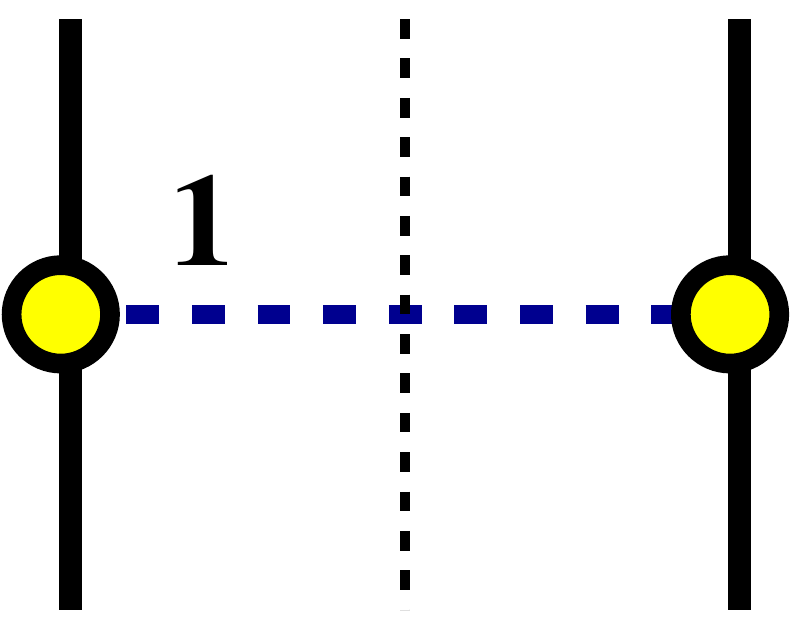}}\;\;
\hbox{\LARGE +}\;\;
2\Re\bigg(
\raisebox{-11pt}{\includegraphics[height=11mm]{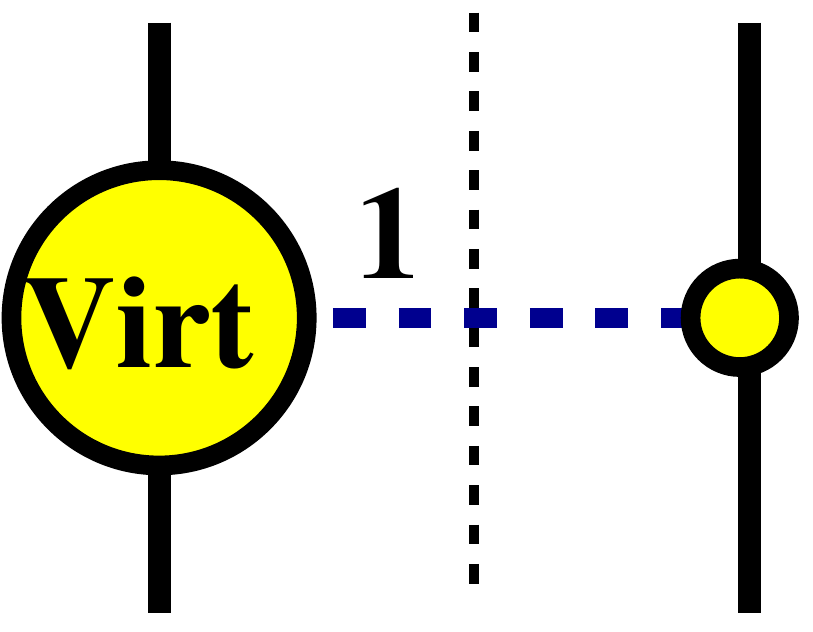}} \bigg)
\hbox{\LARGE +}\;\;
\raisebox{-12pt}{\includegraphics[height=12mm]{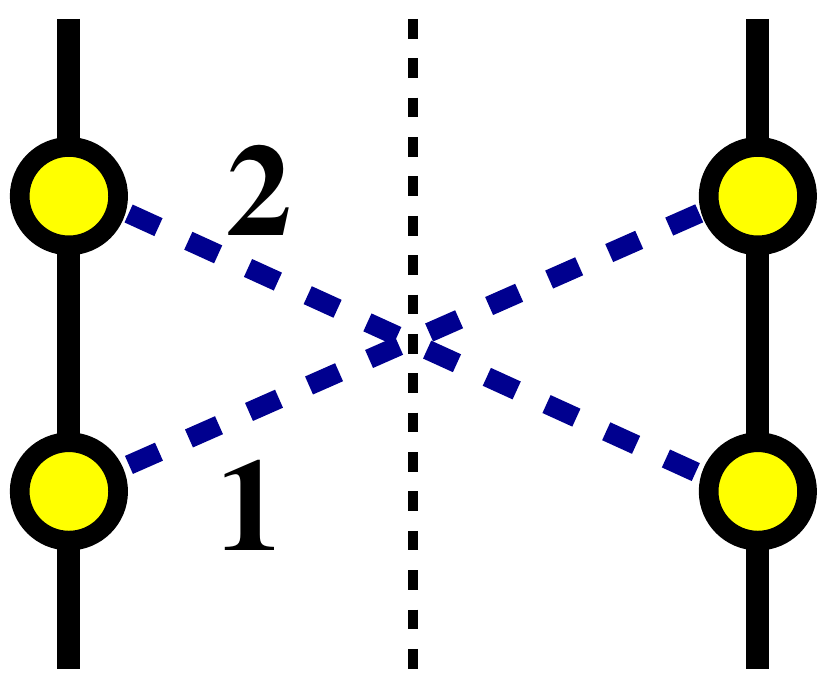}},
\end{equation}
where dashed lines are gluons,
blobs marked ``Virt'' may include several (one loop) subgraphs.
First two terms in the $x$-dependent T.O. exponential with LO kernel read
\begin{equation}
\begin{split}
&\exp_{TO}\left( \Pbbm'_Q\{ K^{LO} \} \right)_x \simeq
\delta_{x=1} +\Pbbm'_Q\{ K_0^{LO} \}_x
=\hbox{\large $\delta_{x=1}+$}
2\Re\bigg(
\raisebox{-11pt}{\includegraphics[height=11mm]{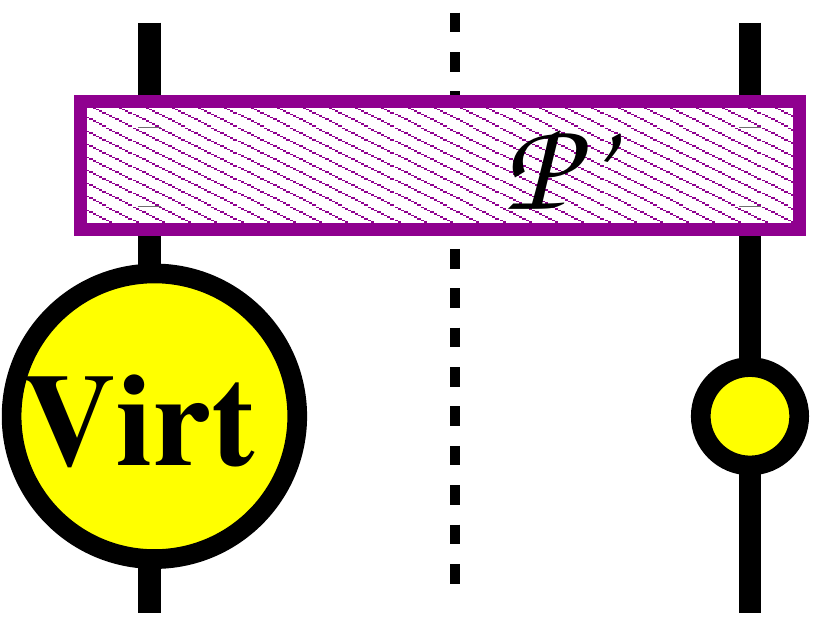}} \bigg)_x
\hbox{\Large +}
\bigg(
\raisebox{-11pt}{\includegraphics[height=11mm]{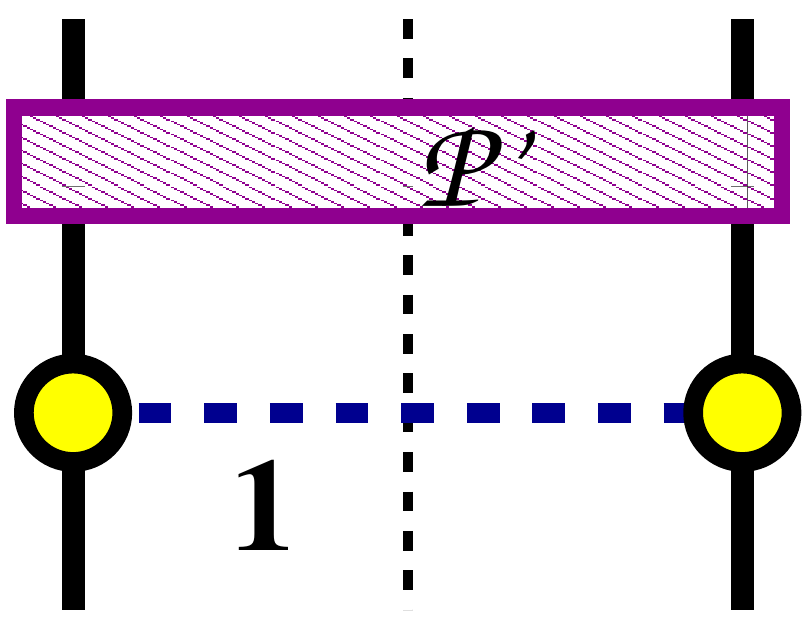}}\bigg)_x
\\&~~~~~~~~
=\delta_{x=1}
+\ln\frac{Q}{q_0}\;
\frac{2C_F\alpha_s}{\pi} \left(\frac{1+x^2}{2(1-x)}\right)_+
=\delta_{x=1} + \ln\frac{Q}{q_0}\; \Peu_{qq}(x),
\end{split}
\end{equation}
where $q_0$ is IR cut-off and
$\Peu_{qq}(x)= \frac{2C_F\alpha_s}{\pi} \left(\frac{1+x^2}{2(1-x)}\right)_+$.

The LO exclusive distribution%
\footnote{Here $\delta$ is another IR cut-off parameter, the same as in CFP.}
$\bar\rho_{B1r}=
\frac{2C_F\alpha_s}{\pi^2}\;
\frac{1+(1-\alpha_1)^2}{2}\;
\frac{1}{\bk_1^2 }
\theta_{\alpha_1>\delta}
$
resides inside
\begin{equation}
\begin{split}
&\raisebox{-11pt}{\includegraphics[height=11mm]{xBrem1rP.pdf}}
=\int \frac{d^3 k_1}{2k_1^0}\; \theta_{Q>a_1>q_0}\;
  \bar\rho_{B1r}(k_1)
=\int \alpha_1 d\alpha_1 d^2 \ba_1 d\phi_1\;
  \theta_{Q>a_1>q_0}\;
  \bar\rho_{B1r}(k_1),
\end{split}
\end{equation}
where
$\ba_i \equiv \bk_1/\alpha_1$ and  $a_1 = |\ba_1|$ $\simeq$ polar angle
of the gluon with transverse momentum $\bk_1$.
The trivial phase space integration gives Sudakov double log or LO kernel:
\begin{equation}
\raisebox{-11pt}{\includegraphics[height=11mm]{xBrem1rP.pdf}}
=\frac{2C_F\alpha_s}{\pi} \ln\frac{Q}{q_0}
  \left( \ln\frac{1}{\delta} -\frac{3}{4} \right)
= S_{_{\rm ISR}},\quad \quad \quad
\bigg(
\raisebox{-11pt}{\includegraphics[height=11mm]{xBrem1rP.pdf}}\bigg)_x
=\ln\frac{Q}{q_0}\; \Peu_{qq}(x) \theta_{1-x<\epsilon}.
\end{equation}
In the TO exponential
$D(Q)=\exp_{TO}\big( \Pbbm'_Q\{ K^{LO} \} \big)$
the virtual contribution
~\raisebox{-8pt}{\includegraphics[height=8mm]{xBrem1vP.pdf}}$=-S_{ISR}$
factorizes off:
\begin{equation}
\begin{split}
&D(Q)\!
=1 +\Pbbm'_Q\{ K_0 \}
   +\Pbbm'_Q\{^{s_2}   K_0 \}\cdot\Pbbm'_{s_2}\{^{s_1} K_0 \}
   +\Pbbm'_Q\{^{s_3}   K_0 \}\cdot
    \Pbbm'_{s_3}\{^{s_2} K_0 \}\cdot\Pbbm'_{s_2}\{^{s_1} K_0 \}
   +...
\\&
=\exp\left(
\Re\;\raisebox{-10pt}{\includegraphics[height=10mm]{xBrem1vP.pdf}}
\right)\;
\left\{
  \raisebox{-10pt}{\includegraphics[height=10mm]{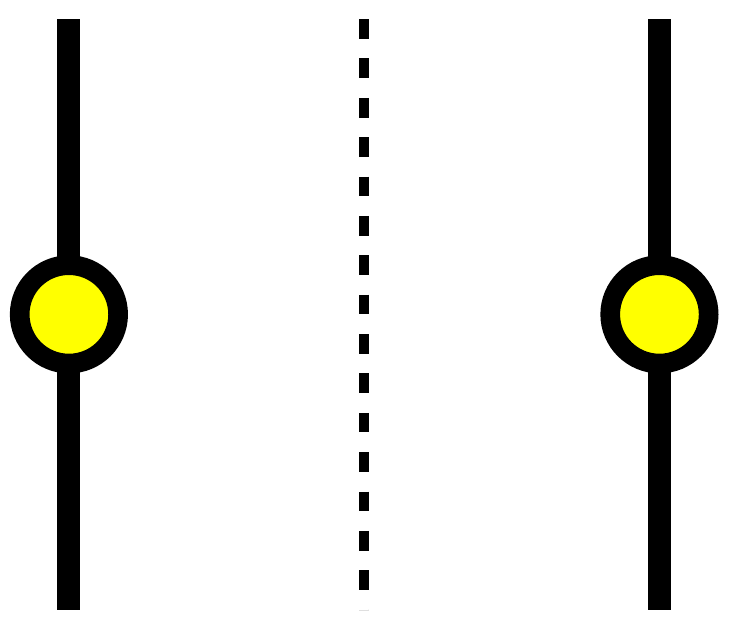}}
  \hbox{\LARGE +}\;\;
  \raisebox{-10pt}{\includegraphics[height=10mm]{xBrem1rP.pdf}}
  \hbox{\LARGE +}\;\;
  \raisebox{-10pt}{\includegraphics[height=10mm]{xB2P1P}}
  \hbox{\LARGE +}\;\;
  \raisebox{-12pt}{\includegraphics[height=11mm]{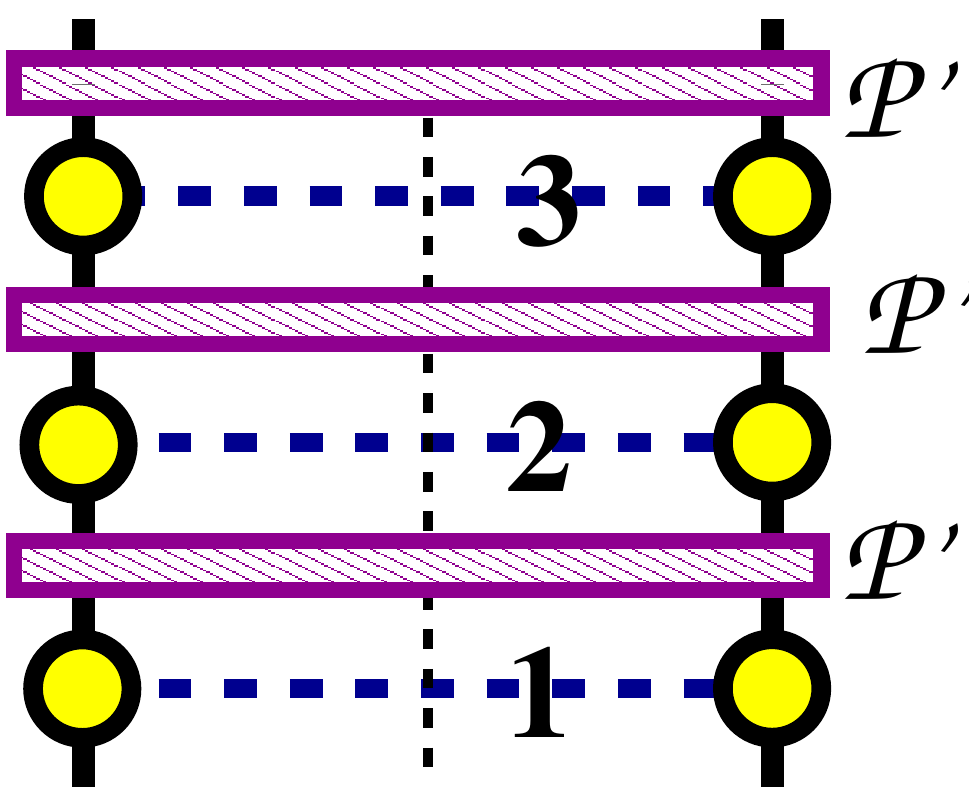}}
  \hbox{\LARGE +...}
\right\}
\\&
=e^{-S_{_{\rm ISR}}}\bigg\{1+
\sum_{n=1}^\infty
\prod_{j=1}^n
  \int \frac{d^3 k_j}{2k_j^0}\; \Pi(k_j)\;
  \bar\rho_{B1}^{LO}(k_j)
  \theta_{a_j>a_{j-1}}
\bigg\}
=e^{-S_{_{\rm ISR}}}\; \left\{ e^{+S_{_{\rm ISR}}}\right\}=1.
\end{split}
\end{equation}
The $x$-dependent version of the same 
$D(Q,x)=\exp_{TO}\bigg( \Pbbm'_Q\{ K^{LO} \} \bigg)_x$ reads
\begin{equation}
D(x,Q)
=e^{-S_{_{\rm ISR}}}\bigg\{\delta_{x=1}+
\sum_{n=1}^\infty
\bigg(\prod_{j=1}^n\;
  \int_{q_{j-1}}^Q \frac{d a_j}{a_j}
  \int_0^{1-\delta} dz_j
  \int_0^{2\pi}\frac{d \varphi_j}{2\pi}\;
  \Peu^\theta_{qq}(z_j)
\bigg)\delta_{x=\prod_i z_i}
\bigg\}.
\end{equation}

From now on we enter NLO world. 
The NLO kernel$ K^{NLO}_\mu
  =\Pop'_\mu \left\{^s \big(K_0 +K_0\cdot (1-\Pop')\cdot K_0 \big)\right\}$
with $K_0$ of eq.~(\ref{eq:K0NLO}) (nonsinglet bremsstrahlung)
is inserted into T.O. exponent of the NLO ePFD:
\begin{equation}
\label{eq:ePDFnlo}
\begin{split}
&D^{[1]}_{B}(Q)
= \exp(-S^{[1]}_{_{ISR}})
\Big( 1 +\Pbbm'_Q\{ K^{r} \}
   +\Pbbm'_Q\{^{a_2}   K^{r} \}\cdot\Pbbm'_{a_2}\{^{a_1} K^{r} \}+
\\&~~~~~~~~~~~~~~~~~~~~~~~~~~~~~~~~~~~~~~~~~
   +\Pbbm'_Q\{^{a_3}   K^{r} \}\cdot
    \Pbbm'_{a_3}\{^{a_2} K^{r} \}\cdot\Pbbm'_{a_2}\{^{a_1} K^{r} \}
   +\dots\Big)
\\&~
=\exp\left(
\Re\raisebox{-8pt}{\includegraphics[height=8mm]{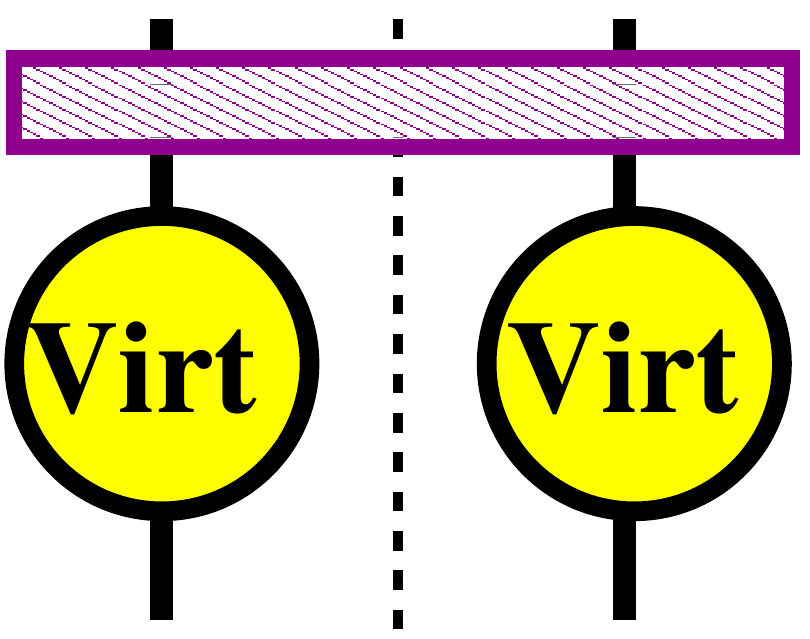}}
\right)
\Bigg\{
  \raisebox{-9pt}{\includegraphics[height=9mm]{xBorn.pdf}}
  \hbox{\LARGE +}
  \raisebox{-9pt}{\includegraphics[height=9mm]{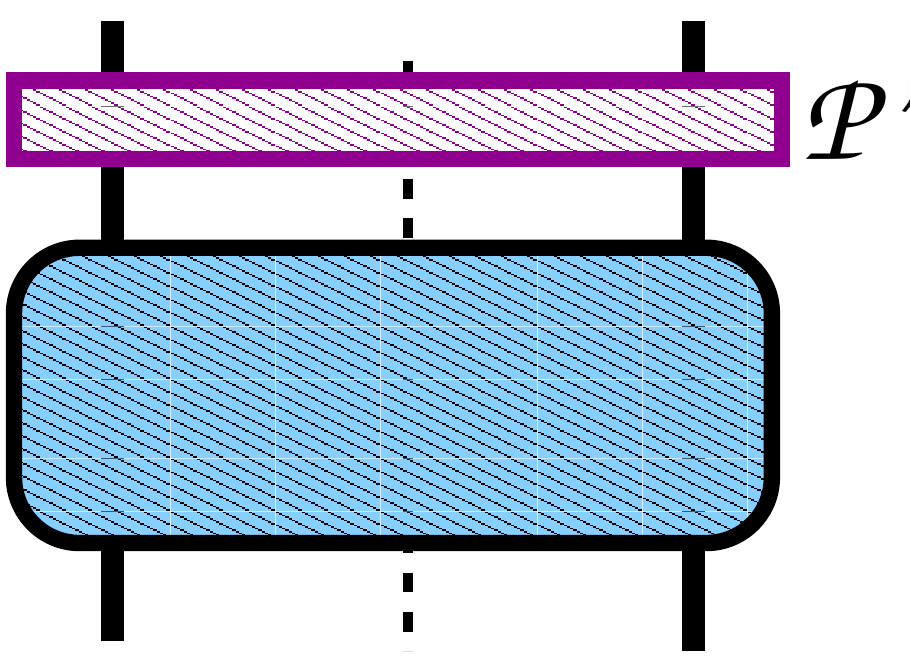}}
  \hbox{\LARGE +}
  \raisebox{-9pt}{\includegraphics[height=9mm]{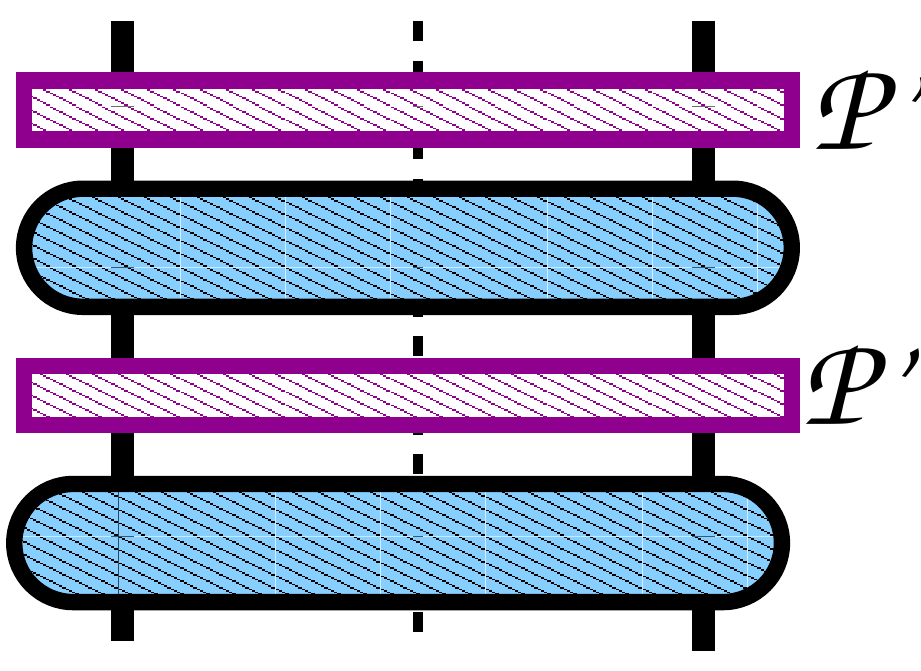}}
  \hbox{\LARGE +}\;\;\;
  \raisebox{-12pt}{\includegraphics[height=12mm]{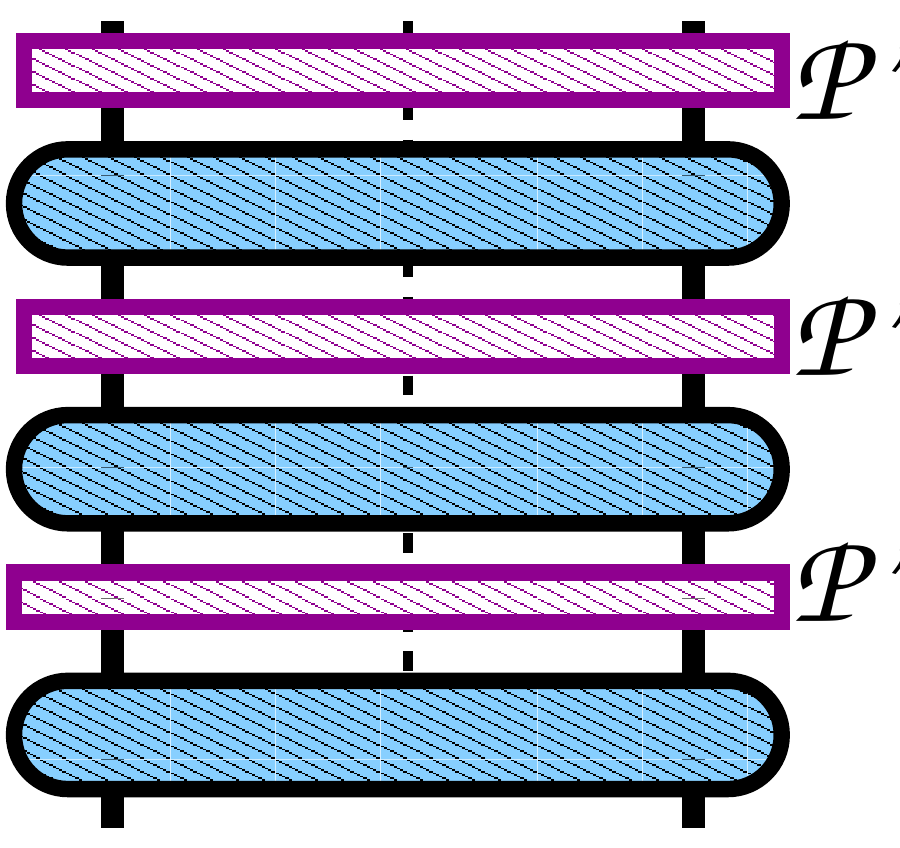}}
  \hbox{\LARGE +...}
\Bigg\}.
\end{split}
\end{equation}
Where the zero-real-emission part of the kernel
$-S^{[1]}_{ISR}=$\raisebox{-8pt}{\includegraphics[height=8mm]{xBrem2vvP.pdf}}
(wave function renormalization up to second order)
factorizes and exponentiates.

The remaining part of the NLO kernel
$ K^r=
\raisebox{-9pt}{\includegraphics[height=9mm]{xKlad1.pdf}}\;
\hbox{\LARGE $\equiv$}\;\;
\raisebox{-9pt}{\includegraphics[height=9mm]{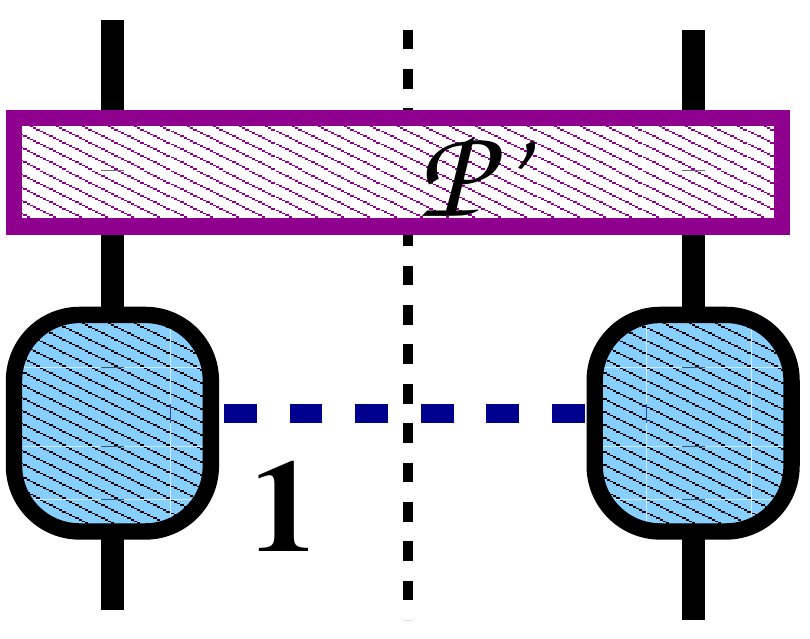}}\;
  \hbox{\LARGE +}\;
\raisebox{-9pt}{\includegraphics[height=9mm]{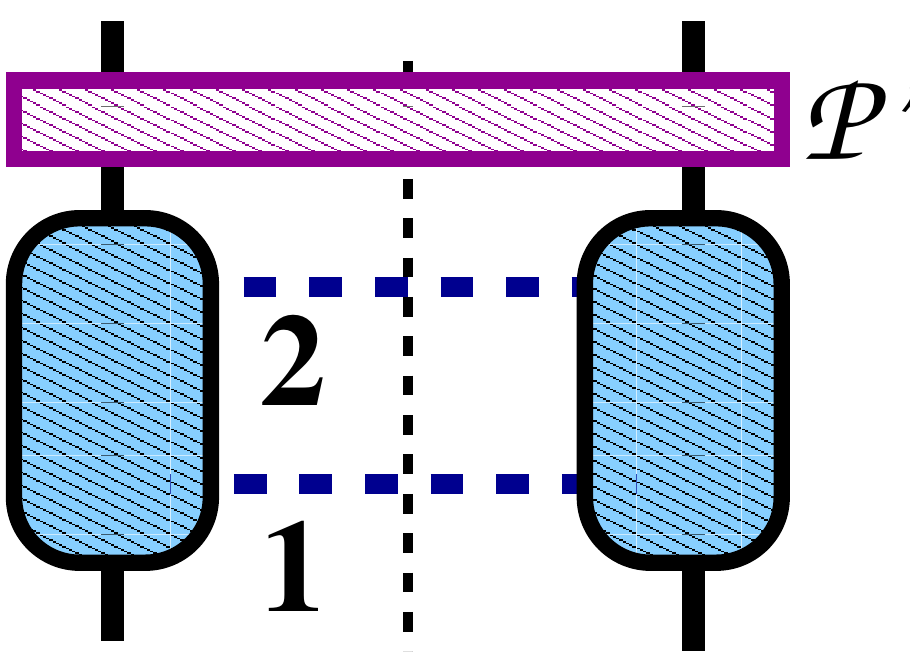}}
$
consists of the one-real-emission part including NLO virtual one loop corrections
$K^{1r}=
\raisebox{-9pt}{\includegraphics[height=9mm]{xKseg1vP.pdf}}\; 
\hbox{\LARGE $\equiv$}\;
\raisebox{-9pt}{\includegraphics[height=9mm]{xBrem1rP.pdf}}\; 
\hbox{\LARGE +}\;\;
2\Re\;
\raisebox{-9pt}{\includegraphics[height=9mm]{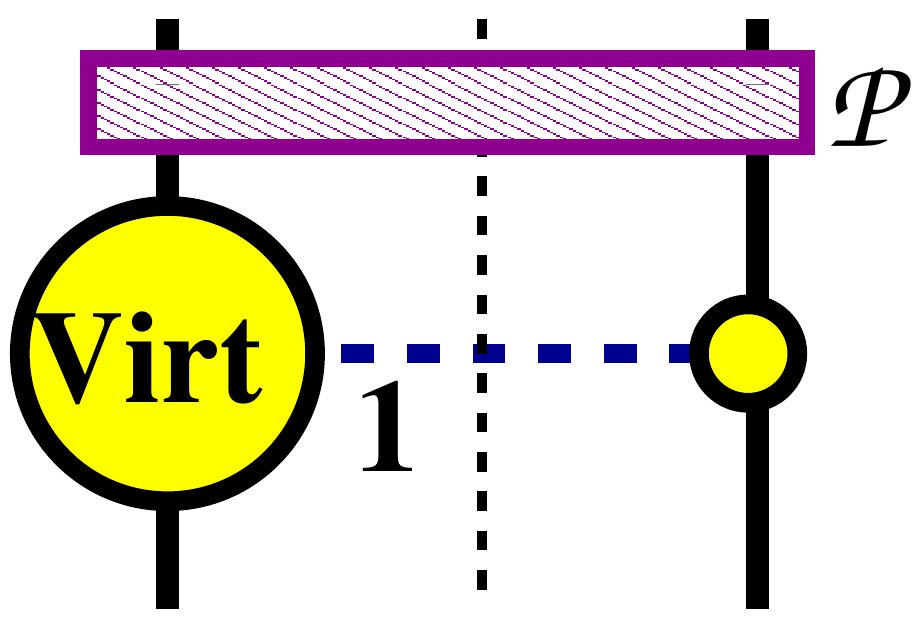}}$
and the pure two-real emission part
$K^{2r}=
\raisebox{-9pt}{\includegraphics[height=9mm]{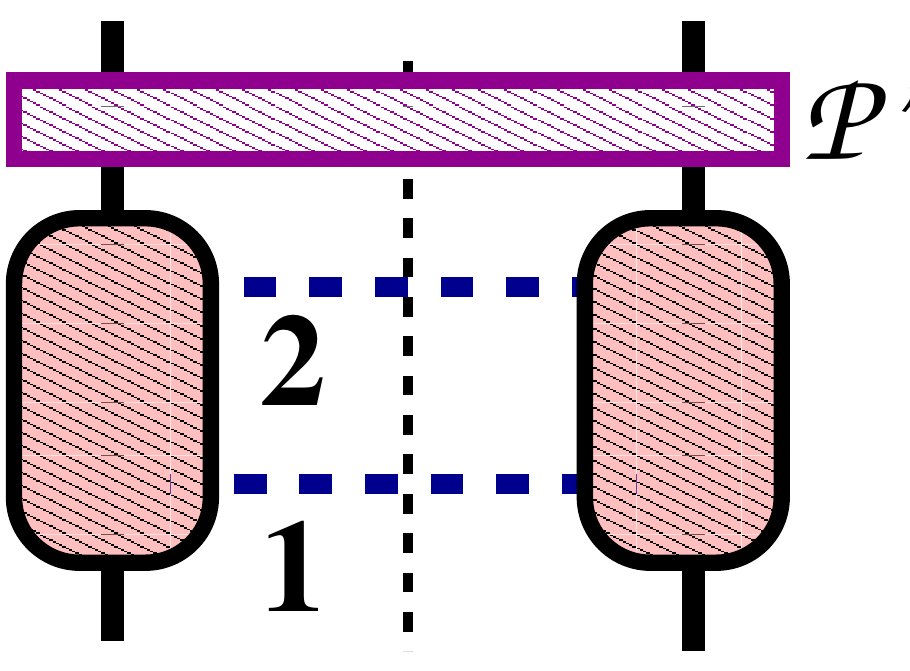}}
\hbox{\LARGE $\equiv$}\;
\raisebox{-9pt}{\includegraphics[height=9mm]{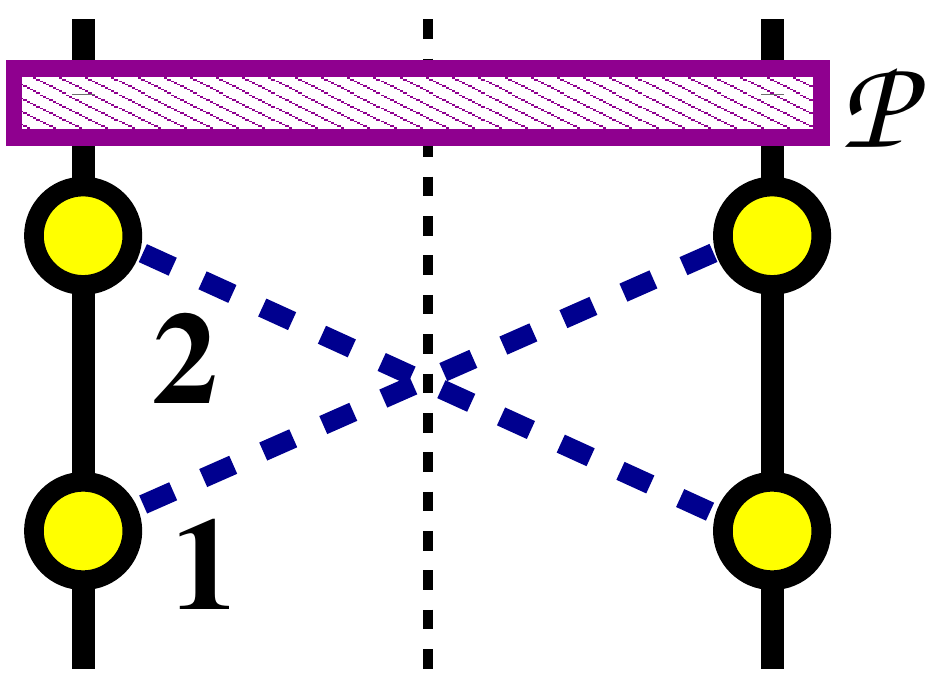}}
\hbox{\LARGE +}\;
\Bigg\{
\raisebox{-9pt}{\includegraphics[height=9mm]{xB21P}}\;
\hbox{\LARGE -}\;
\raisebox{-9pt}{\includegraphics[height=9mm]{xB2P1P}}
\Bigg\}
$.
Eq.~(\ref{eq:ePDFnlo}) could be used directly in the MC,
if both $K^{1r}$ and $K^{2r}$ were positive.
Since $K^{2r}$ is non-positive we have to
recombine $K^{2r}$ with 2 real emission distribution,
that is to put it back where it came from 
(reversing what the factorization did):
\begin{equation}
\label{eq:DBQ}
\begin{split}
&D^{[1]}_{B}(Q)
= e^{-S^{[1]}_{_{ISR}}}
\Bigg\{
\hbox{\large 1+}\;
\raisebox{-7pt}{\includegraphics[height=7mm]{xKseg1vP.pdf}}
\hbox{\Large +}
\raisebox{-8pt}{\includegraphics[height=8mm]{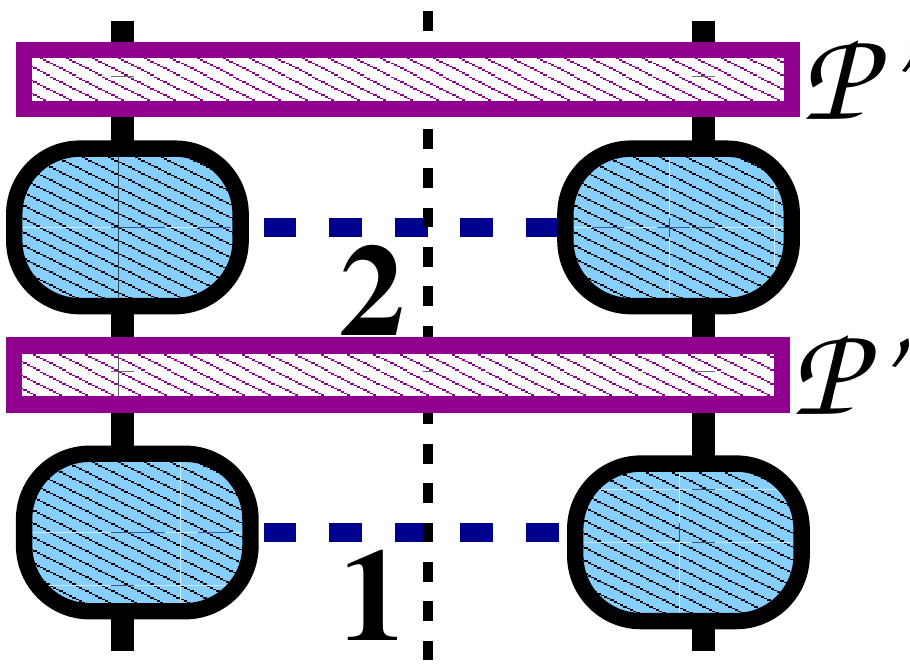}}
\hbox{\Large +}
\raisebox{-9pt}{\includegraphics[height=9mm]{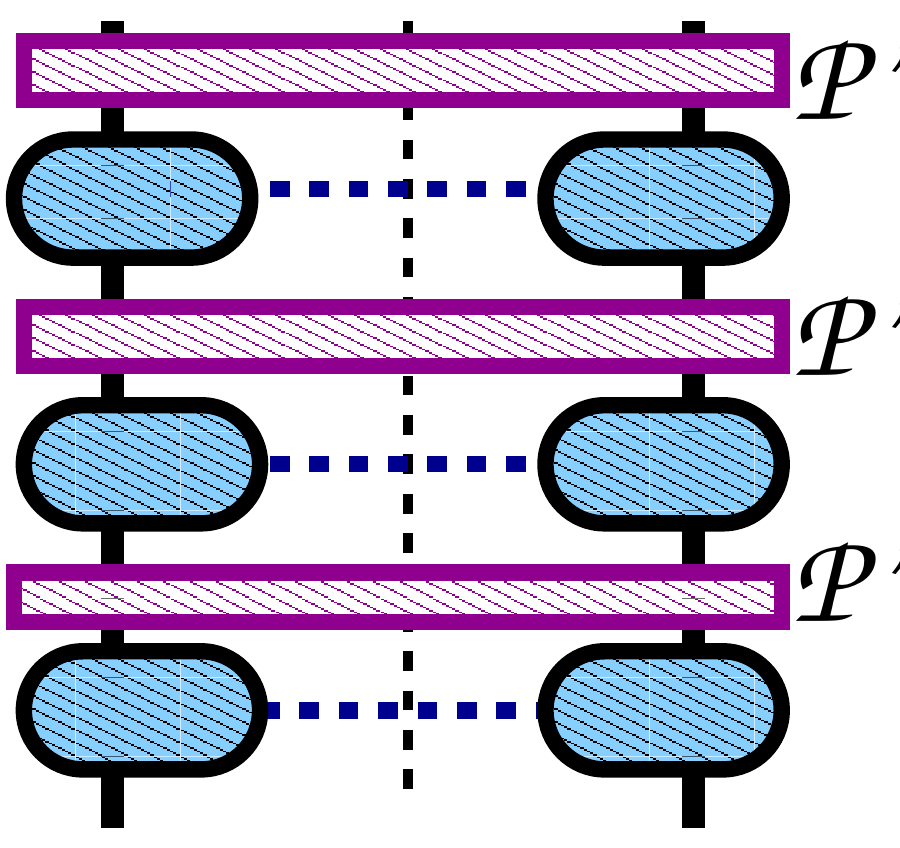}}
\hbox{\Large +}
\raisebox{-11pt}{\includegraphics[height=11mm]{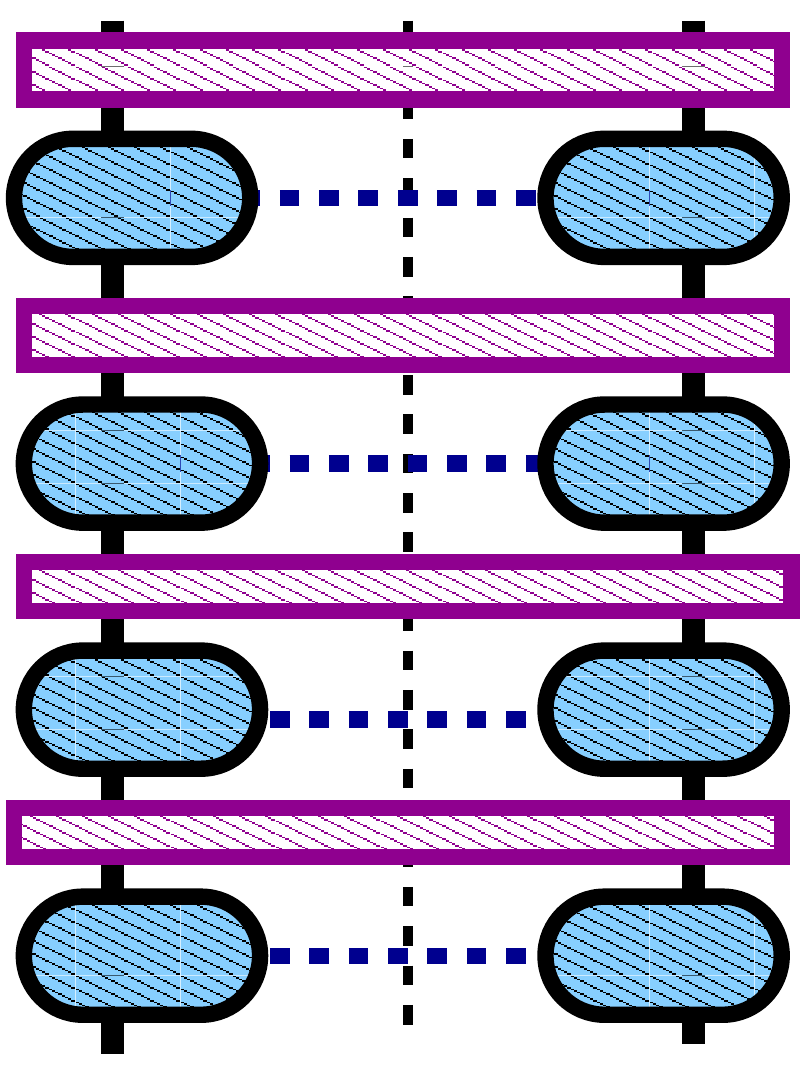}}
\hbox{\Large +...}
\\&~~~~~~~~~~~~~~~~~~~~~~~~~~~~~~~
\hbox{\Large +}
\raisebox{-8pt}{\includegraphics[height=8mm]{xKseg2rPink.pdf}}
\hbox{\Large +}
\raisebox{-9pt}{\includegraphics[height=9mm]{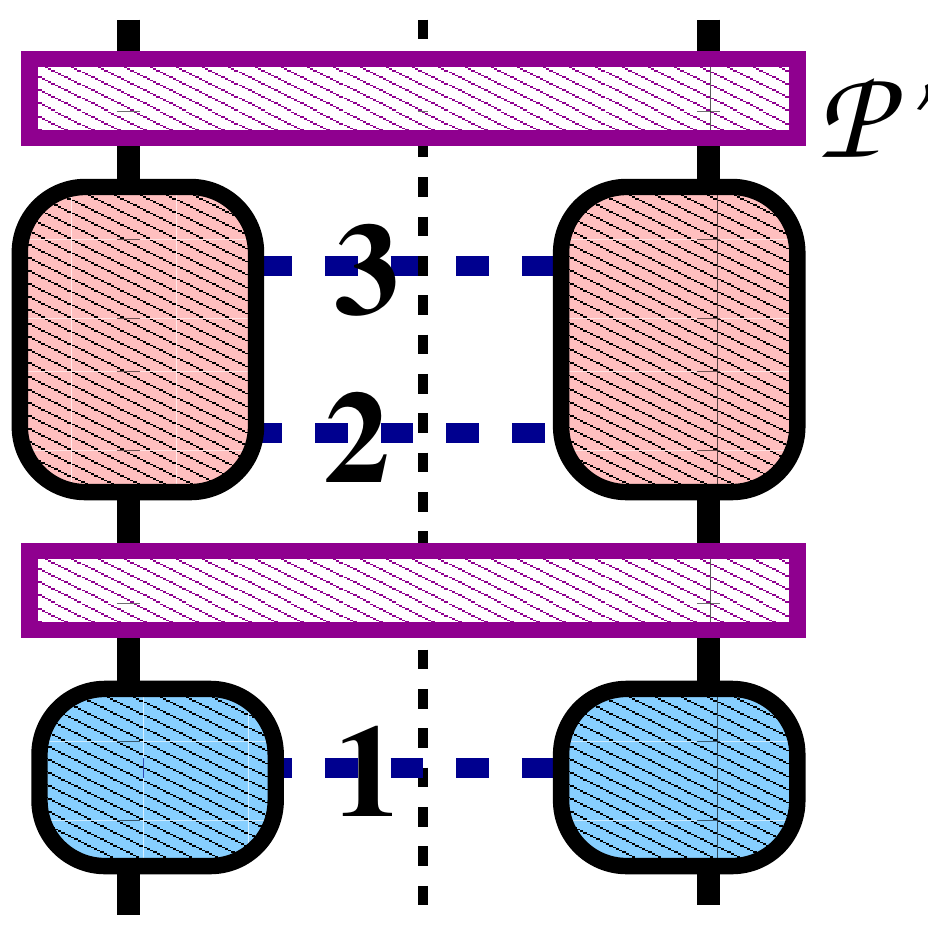}}
\hbox{\Large +}
\raisebox{-9pt}{\includegraphics[height=9mm]{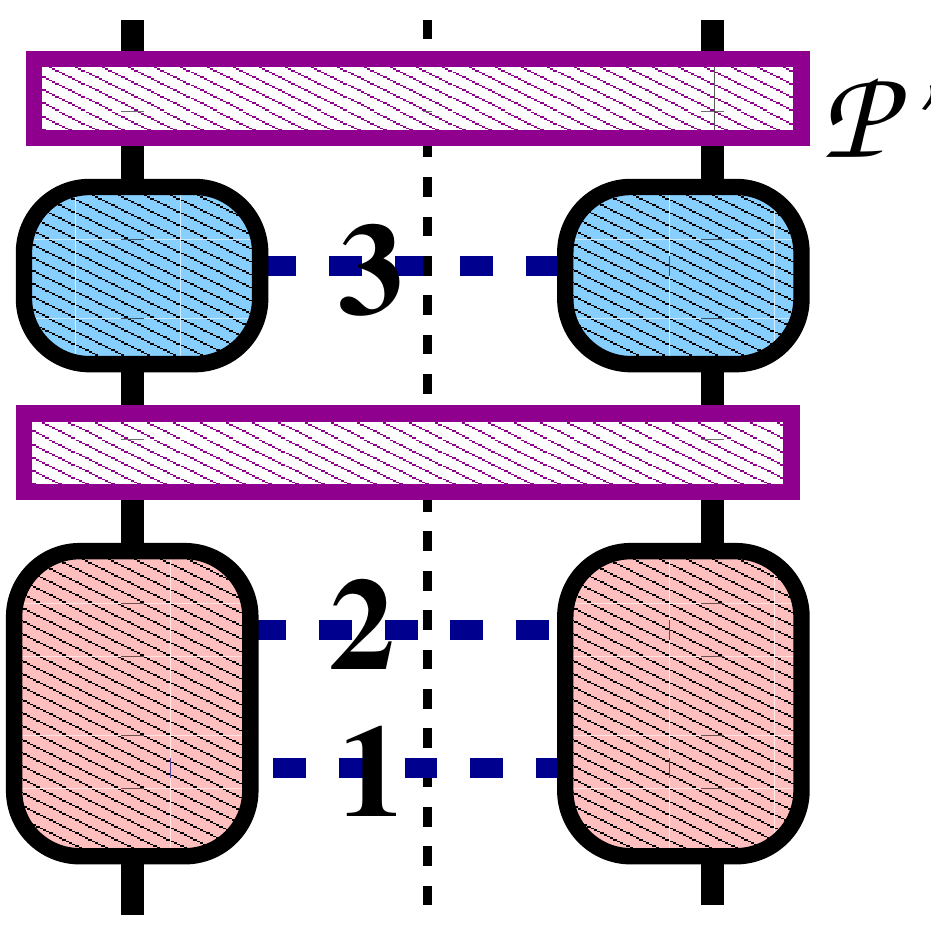}}
\hbox{\Large +}
\raisebox{-11pt}{\includegraphics[height=11mm]{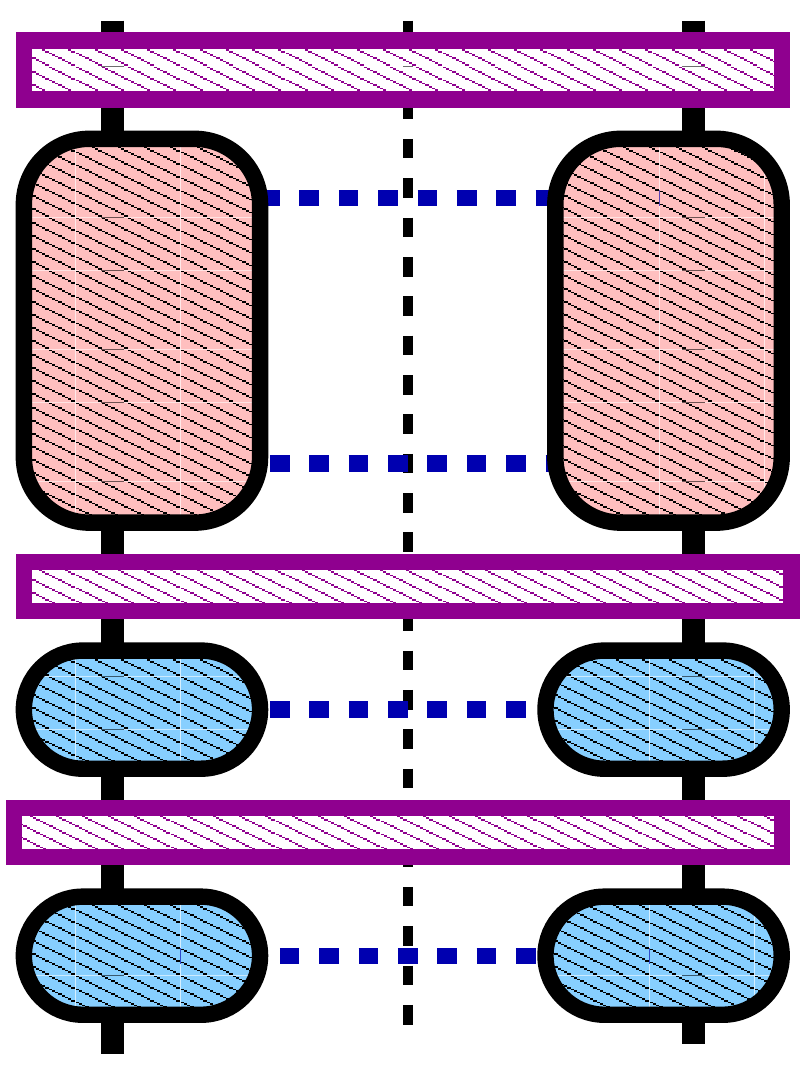}}
\hbox{\Large +...}
\hbox{\Large +}
\raisebox{-11pt}{\includegraphics[height=11mm]{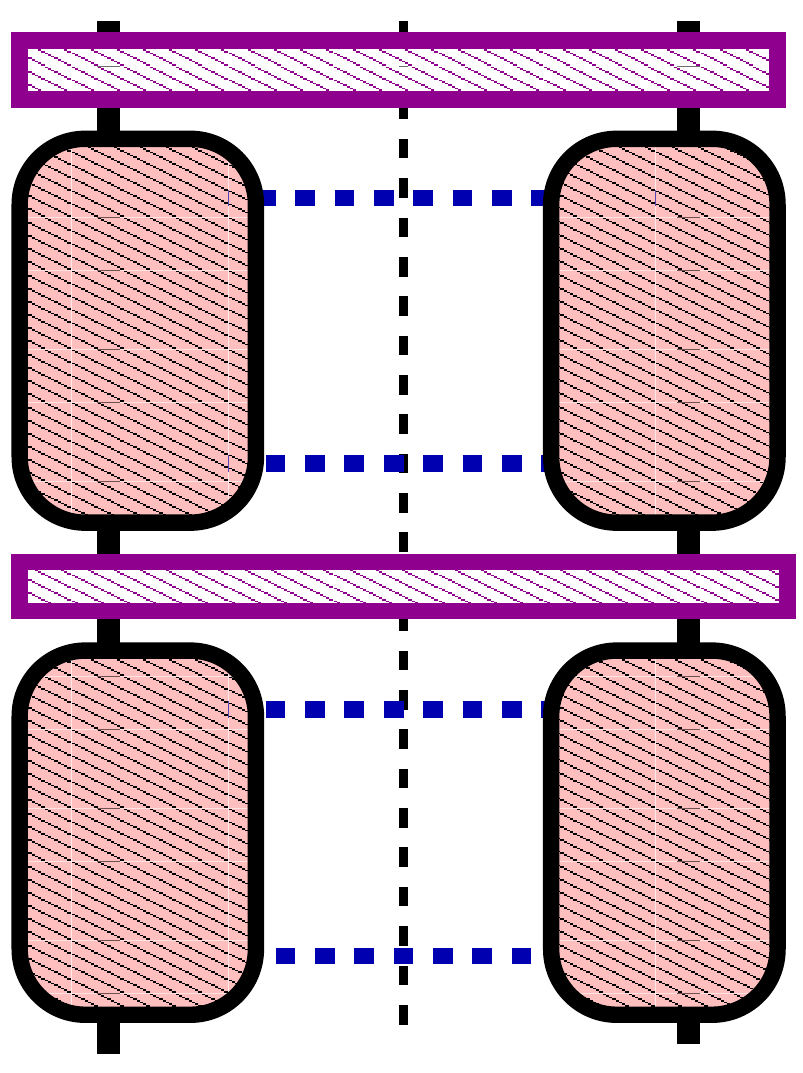}}
\hbox{\Large +...}
\Bigg\}.
\end{split}
\end{equation}
In eq.~(\ref{eq:DBQ})
the first row is positive and large (LO + virtual NLO),
hence it will go to basic MC, while the second row being small and
nonpositive can be absorbed into MC correcting weight.
However, it is not so easy to recombine $K^{2r}$ with 2 real emissions.
The target distribution
\raisebox{-8pt}{\includegraphics[height=8mm]{xKK.pdf}}
\hbox{\Large +}
\raisebox{-8pt}{\includegraphics[height=8mm]{xKseg2rPink.pdf}}
cannot be implemented in the ~(\ref{eq:DBQ}) by means of MC-reweighting
\raisebox{-8pt}{\includegraphics[height=8mm]{xKK.pdf}},
simply because it is zero outside
{\em simplex} $Q>a_2>a_1>q_0$,
while the target distribution is nonzero in the bigger
{\em rectangle} $Q>\max(a_2,a_1)>q_0$.
Going back to original Feynman diagrams we see that the above
problem turns out to be fictitious,
if we properly keep track of the Bose-Einstein (BE) symmetrization:
\begin{equation}
\begin{split}
&
\raisebox{-10pt}{\includegraphics[height=9mm]{xKseg2rPink.pdf}}
\hbox{\LARGE =}\;
2\raisebox{-9pt}{\includegraphics[height=9mm]{xBremXrP.pdf}}
\hbox{\LARGE +}\;
\Bigg\{
\raisebox{-9pt}{\includegraphics[height=9mm]{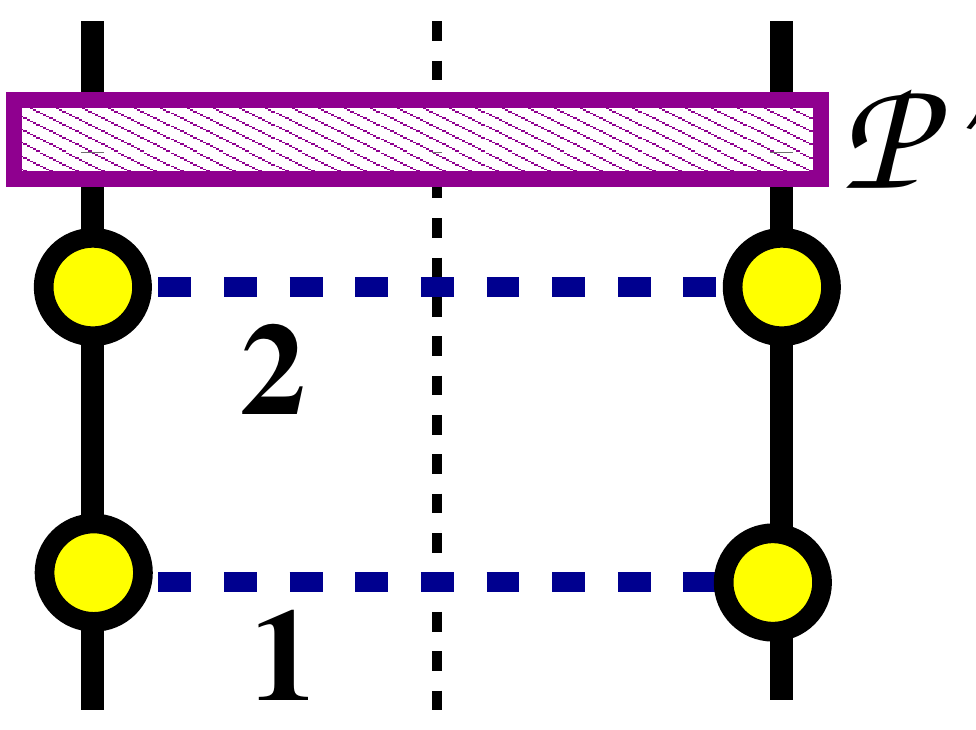}}\;
\hbox{\LARGE -}\;
\raisebox{-9pt}{\includegraphics[height=9mm]{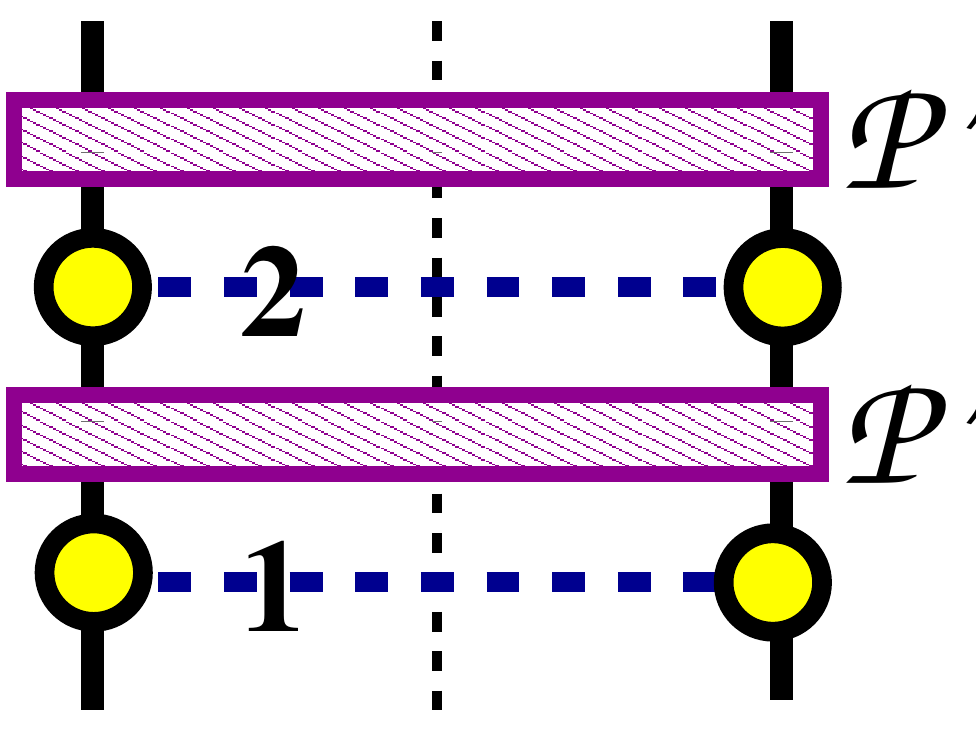}}
\Bigg\}
\hbox{\LARGE +}\;
\Bigg\{
\raisebox{-9pt}{\includegraphics[height=9mm]{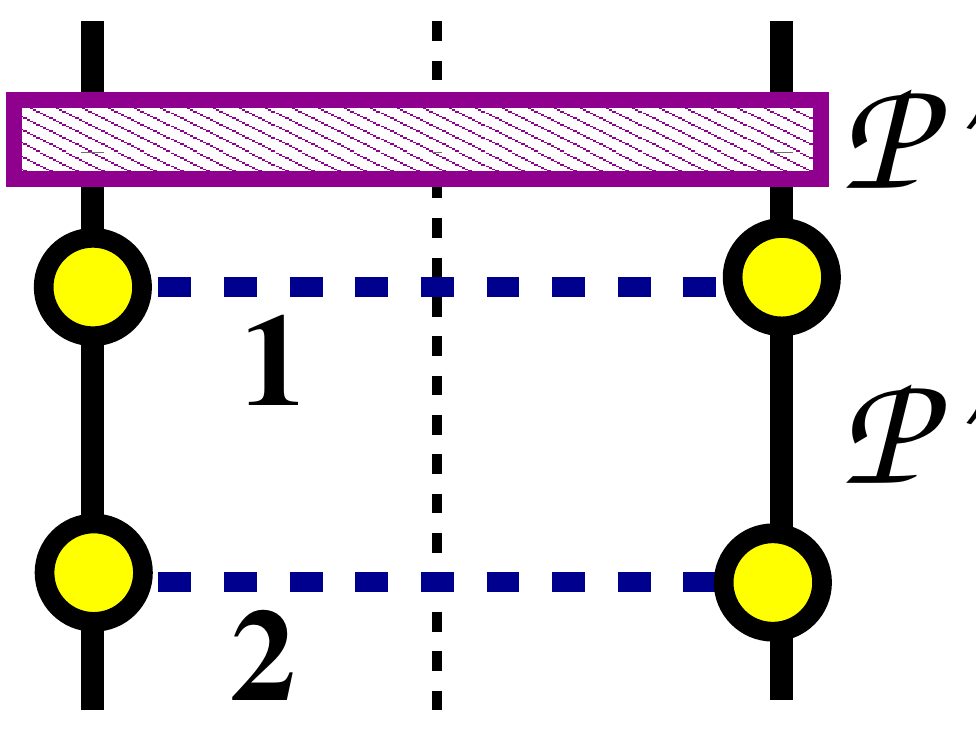}}
\hbox{\LARGE -}\;
\raisebox{-9pt}{\includegraphics[height=9mm]{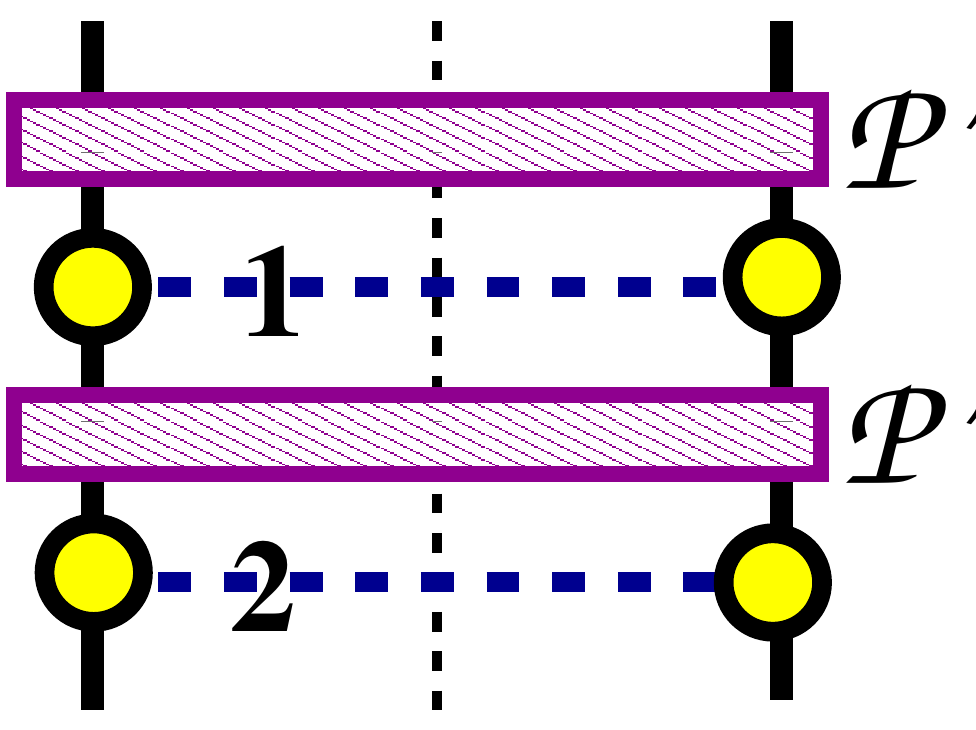}}
\Bigg\}
\\&~~~~~~~~~~~~
= \int \frac{d^3 k_2}{2k_2^0}\;
  \int \frac{d^3 k_1}{2k_1^0}\;
  \theta_{Q>\max(a_2,a_1)>0}\;\theta_{a_2>a_1}
  (\beta_{1B}(k_2,k_1) +\beta_{1B}(k_1,k_2)),
\end{split}
\end{equation}
where for purely technical reasons we include an
internal ordering $\theta_{a_2>a_1}$ for the already symmetric integrand.
See ref.~\cite{Jadach:2009gm} for definition of the NLO 2-gluon function $\beta_{1B}(k_1,k_2)$.
\begin{figure}[!ht]
\centering
\includegraphics[width=80mm,height=40mm]{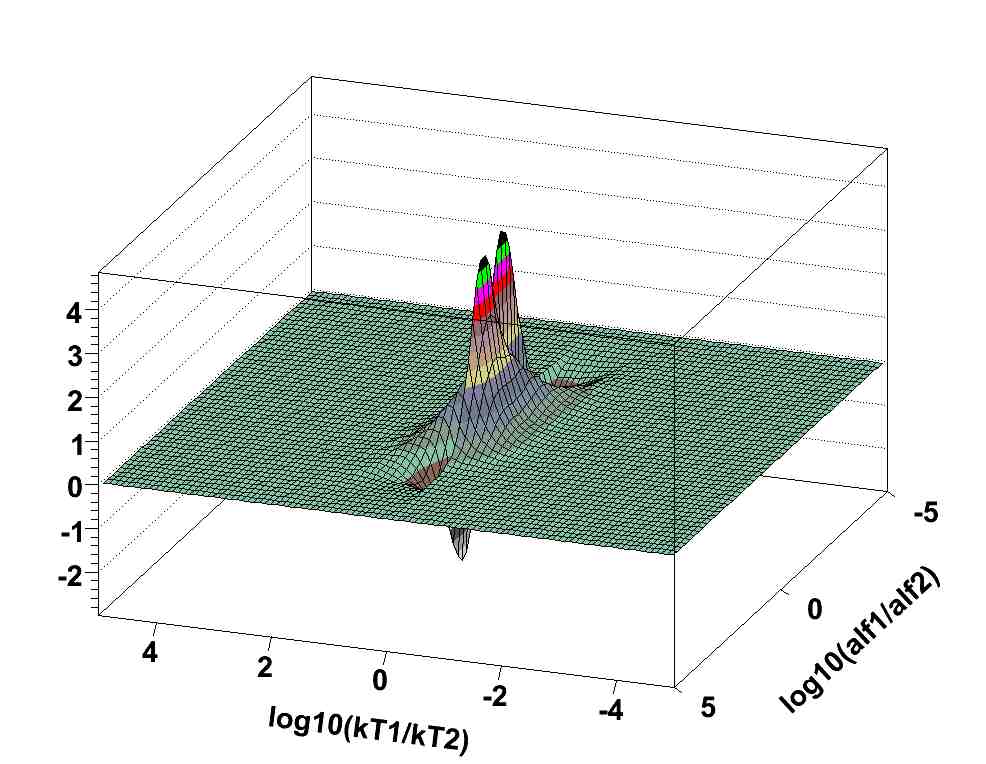}
\caption{
   Two-real-gluon NLO correlation function.
}
\label{fig:two}
\end{figure}
We call sometimes the resulting MC weight the NLO short range {\em correlation} function,
because it contributes  significantly only if both gluons
are non-soft and have transverse momenta (or rapidities) almost equal,
as seen in the plot of Fig.~\ref{fig:two}, see \cite{Jadach:2009gm}.
BE symmetrization requires clever reorganization of combinatorics,
if want to gain on the computation speed by means of
excluding terms equal zero from the BE symmetrization sum.
For instance,
BE symmetrization over $3!$ permutations of 3 arguments of a single NLO correlation function and one
LO spectator distribution reduces to only 2 terms%
\footnote{For simplicity we are drawing here only of half of the ladder.}:
\begin{equation}
\hbox{\LARGE $ \sum\limits_{\{\pi\}} $}\;\;
\raisebox{-10mm}{\includegraphics[height=20mm]{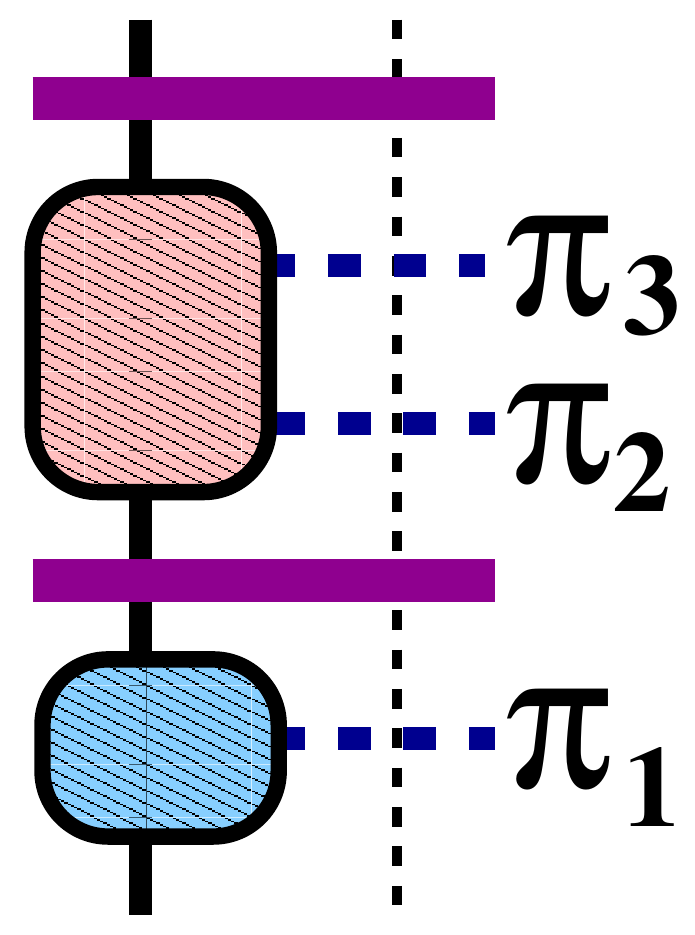}}\;
\hbox{\LARGE $= \sum\limits_{\{\pi^\bullet\}} $}\;\;
\raisebox{-10mm}{\includegraphics[height=20mm]{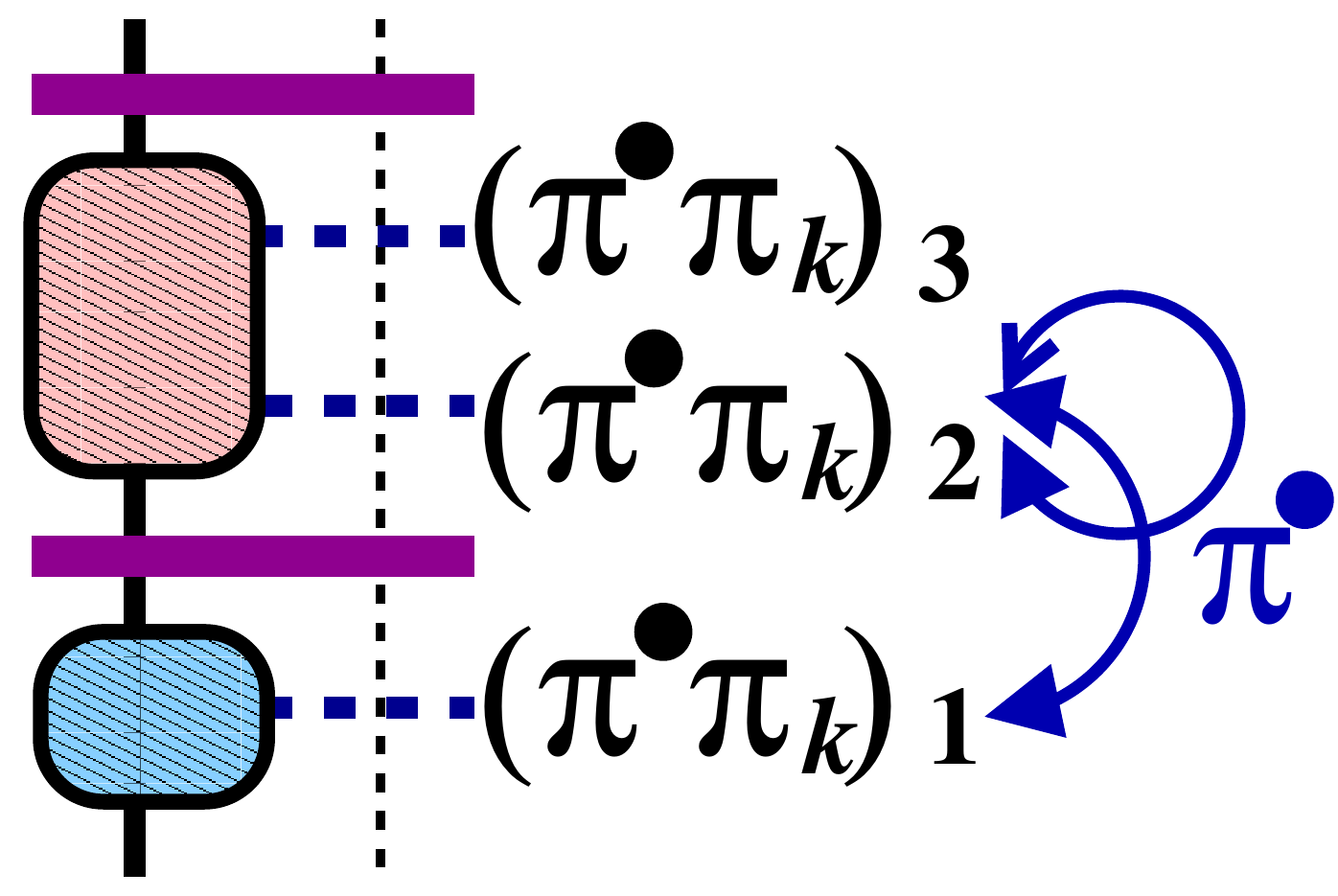}}\;,
\end{equation}
Permutation $\pi_k$ performs the right ordering
$a_{(\pi_k)_3}>a_{(\pi_k)_2}>a_{(\pi_k)_3}$
at a given phase space point $k=(k_1,k_2,k_3)$.
This is denoted in the above by $\{\pi^\bullet \}=\{(123),(213)\}$,
where $(123)$ is identity and $(213)$
interchanges $(\pi_k)_1$ and $(\pi_k)_2$.
Generalization to double insertion of NLO correlation function
in the ladder in a sketchy graphical form is:
\begin{equation}
\hbox{\Large
$\sum\limits_{j_a,j_b}
  \sum\limits_{\{\pi_a^\bullet\}}
  \sum\limits_{\{\pi_b^\bullet\}}
  $}\;
\raisebox{-11mm}{\includegraphics[height=20mm]{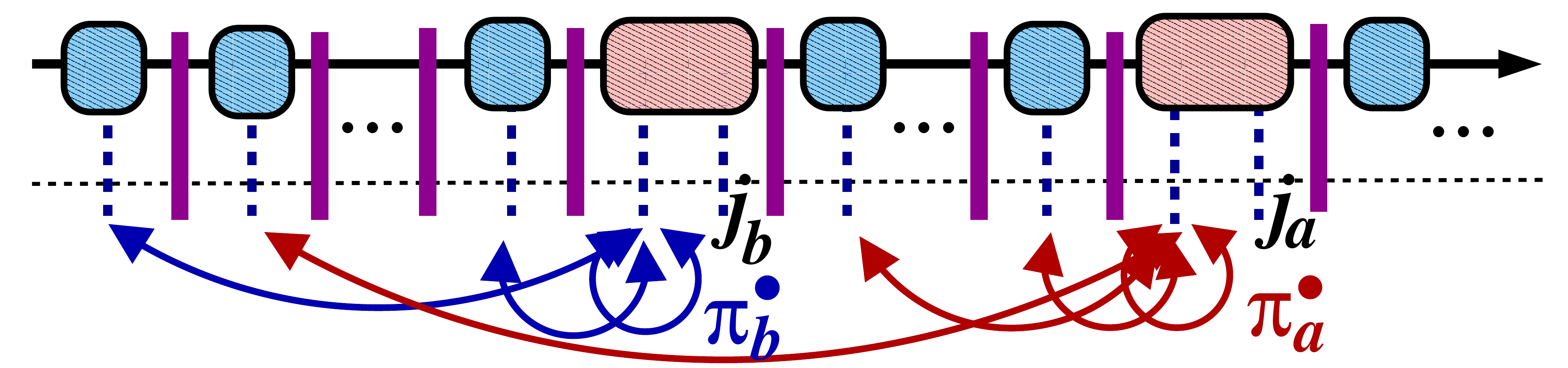}}.
\end{equation}
The corresponding algebraic formula of the above NLO ePDF
can be found in ref.~\cite{Jadach:2009gm}, together with
the numerical result of the precision MC test.

\begin{figure}[!ht]
\centering
\includegraphics[width=100mm,height=60mm]{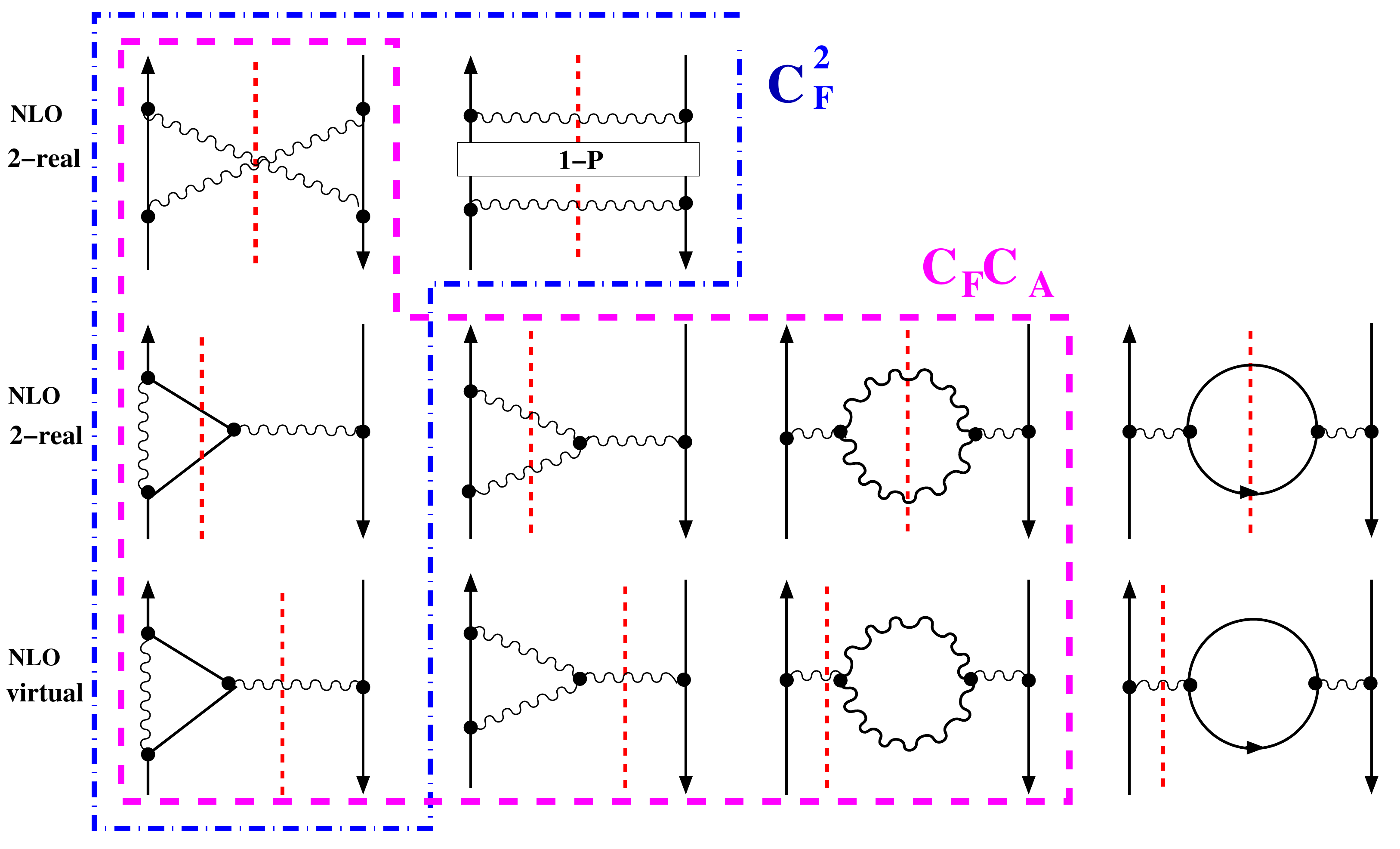}
\caption{
  Feynman diagrams contributing to NLO non-singlet evolution.
}
\label{fig:graphs}
\end{figure}

So far we have considered the $\sim C_F^2$ diagrams, 
in Fig.~\ref{fig:graphs}. 
The remaining 
$\sim C_FC_A$ diagrams add new problem: strong cancellations
between real and virtual contributions
in the NLO correction%
\footnote{ Soft limit of this object was analyzed
  in ref.~\cite{Slawinska:2009gn}, exposing colour coherence effects.}
due to final state radiation (FSR) Sudakov double log,
This enforces exponentiation of the FSR already in the LO basic MC,
if we aim at positive weight MC events.
In the basic MC each gluon in the LO ladder is replaced
by resolved  multigluon
\raisebox{-12pt}{\includegraphics[height=12mm]{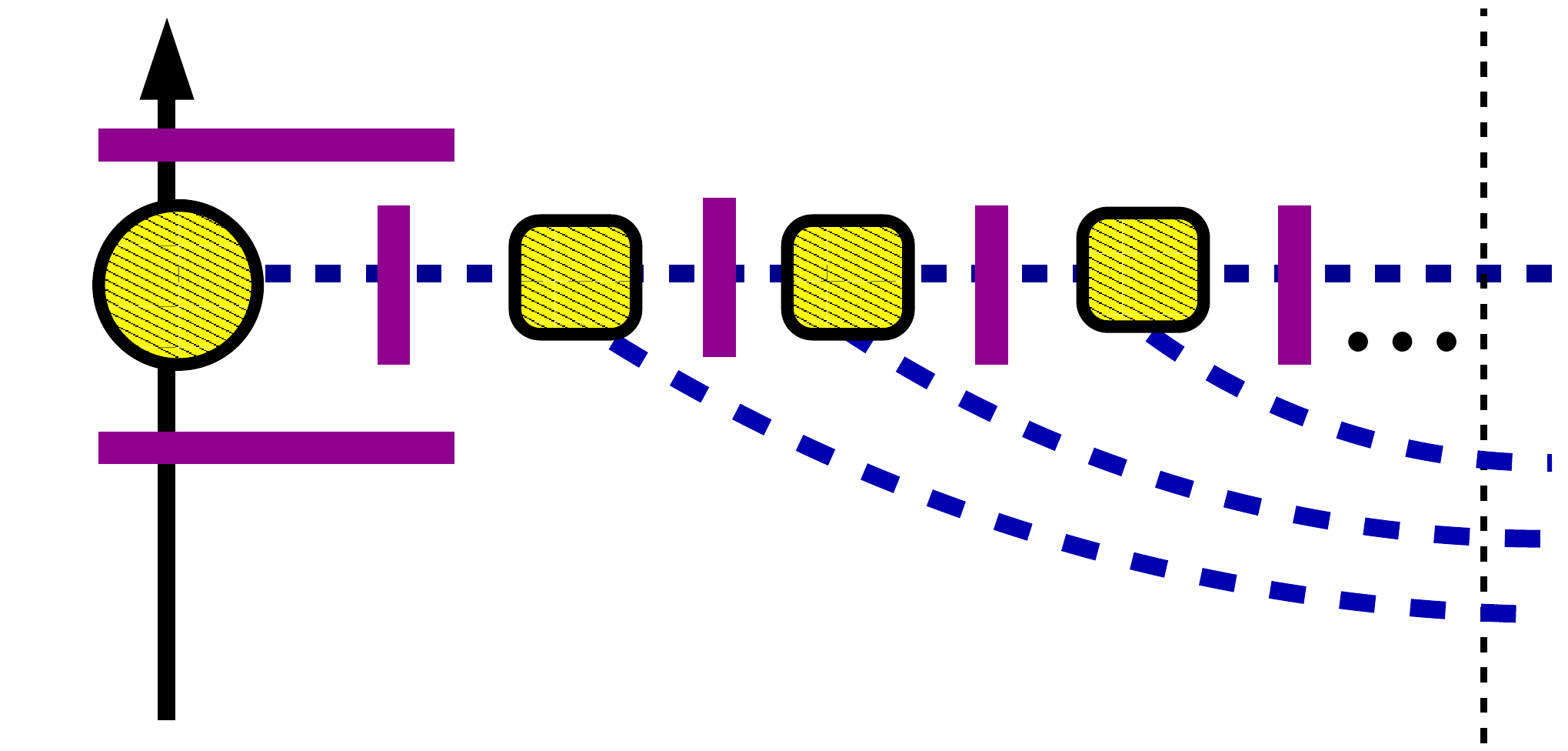}},
obtained by iterating/exponentiating ``soft counterterm''  
\raisebox{-10pt}{\includegraphics[height=10mm]{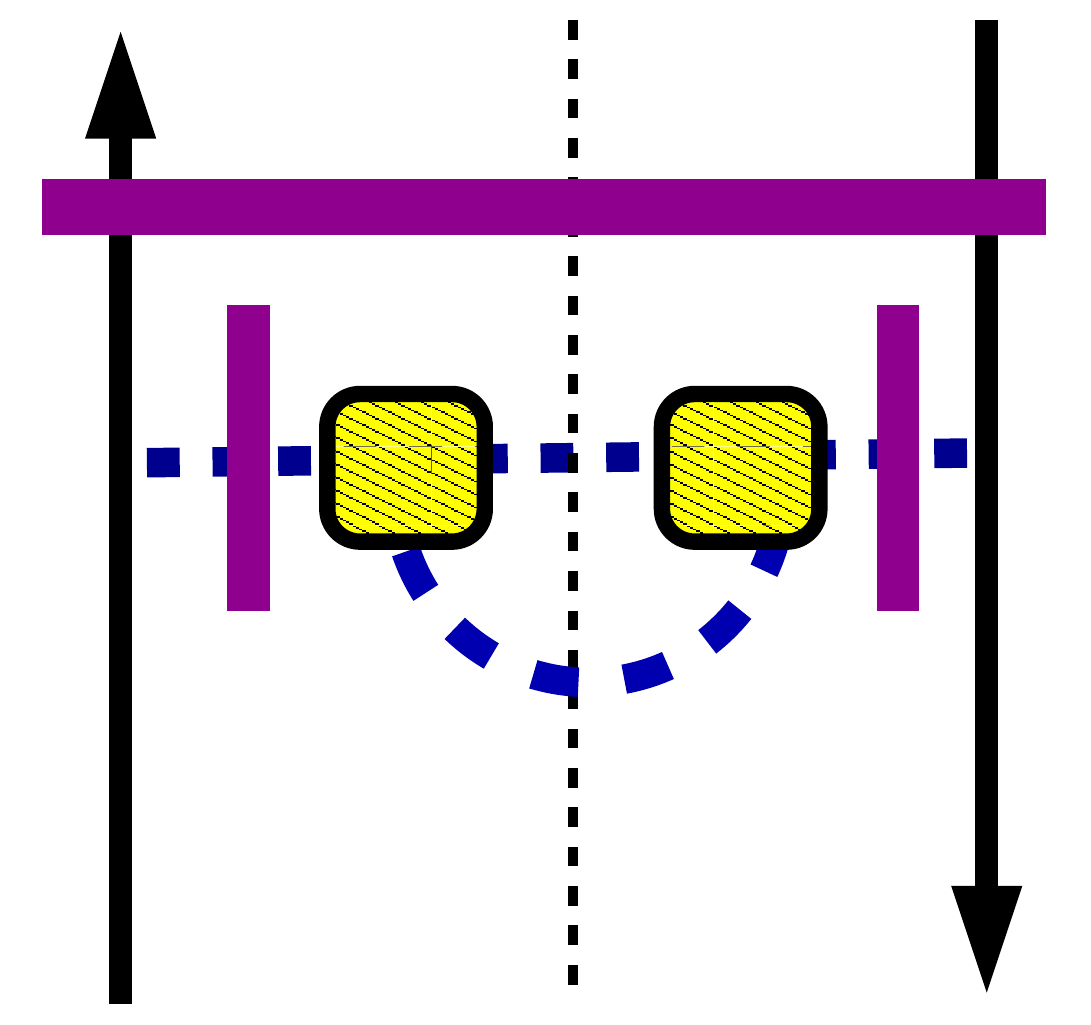}}~in $K^f$.
BE symmetrization is also done for FSR gluons,
such that the complete NLO correcting weight includes the following
sum:
\begin{equation}
\hbox{\large $C_F^2$}\!\!
\hbox{\LARGE $ \sum\limits_{\{\pi^\bullet\}} $}~
\raisebox{-18mm}{\includegraphics[height=37mm]{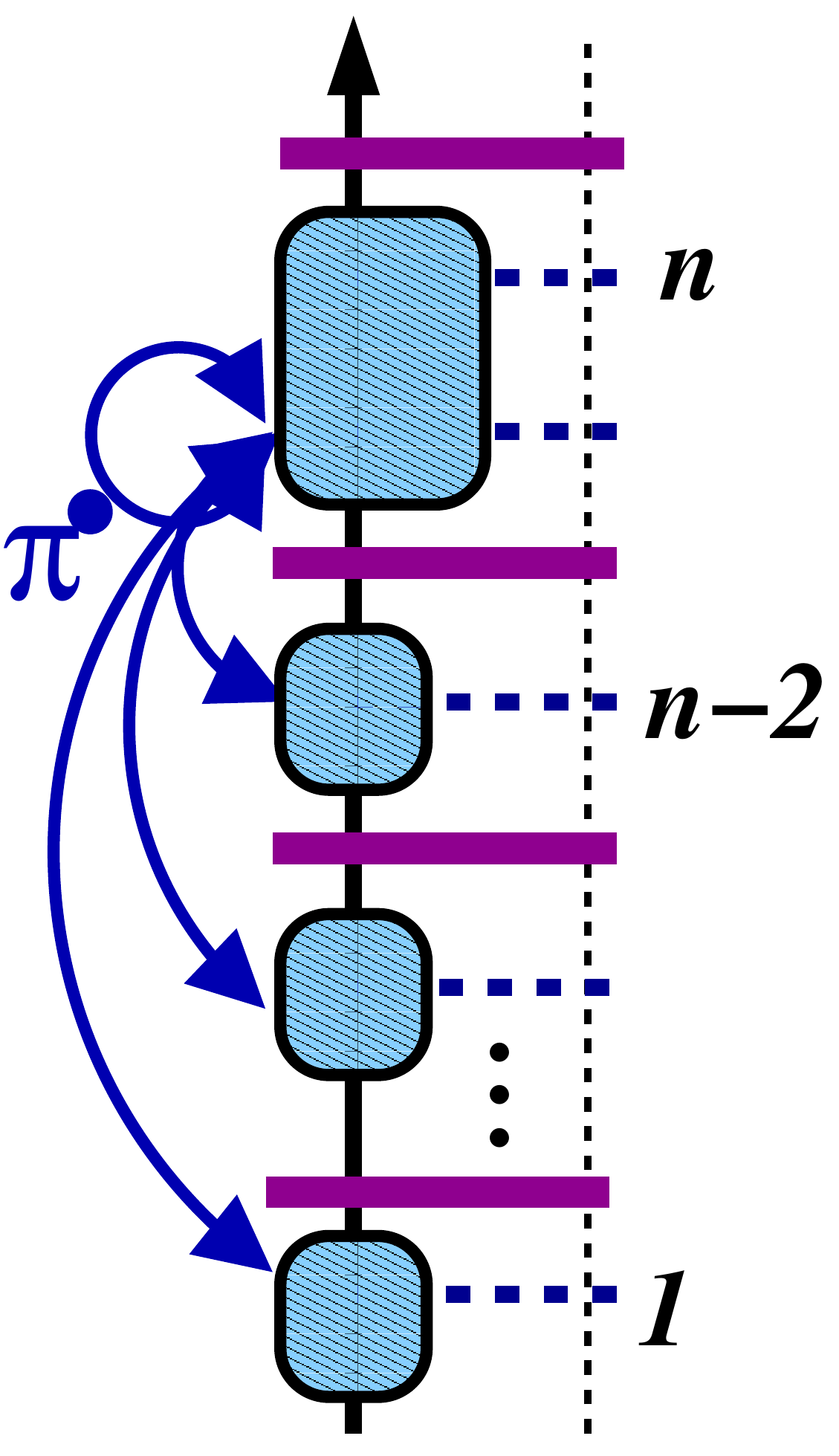}}
\hbox{\large $+C_F C_A$}\!\!
\hbox{\LARGE $ \sum\limits_{\{\pi^\star \}} $}~
\raisebox{-18mm}{\includegraphics[height=37mm]{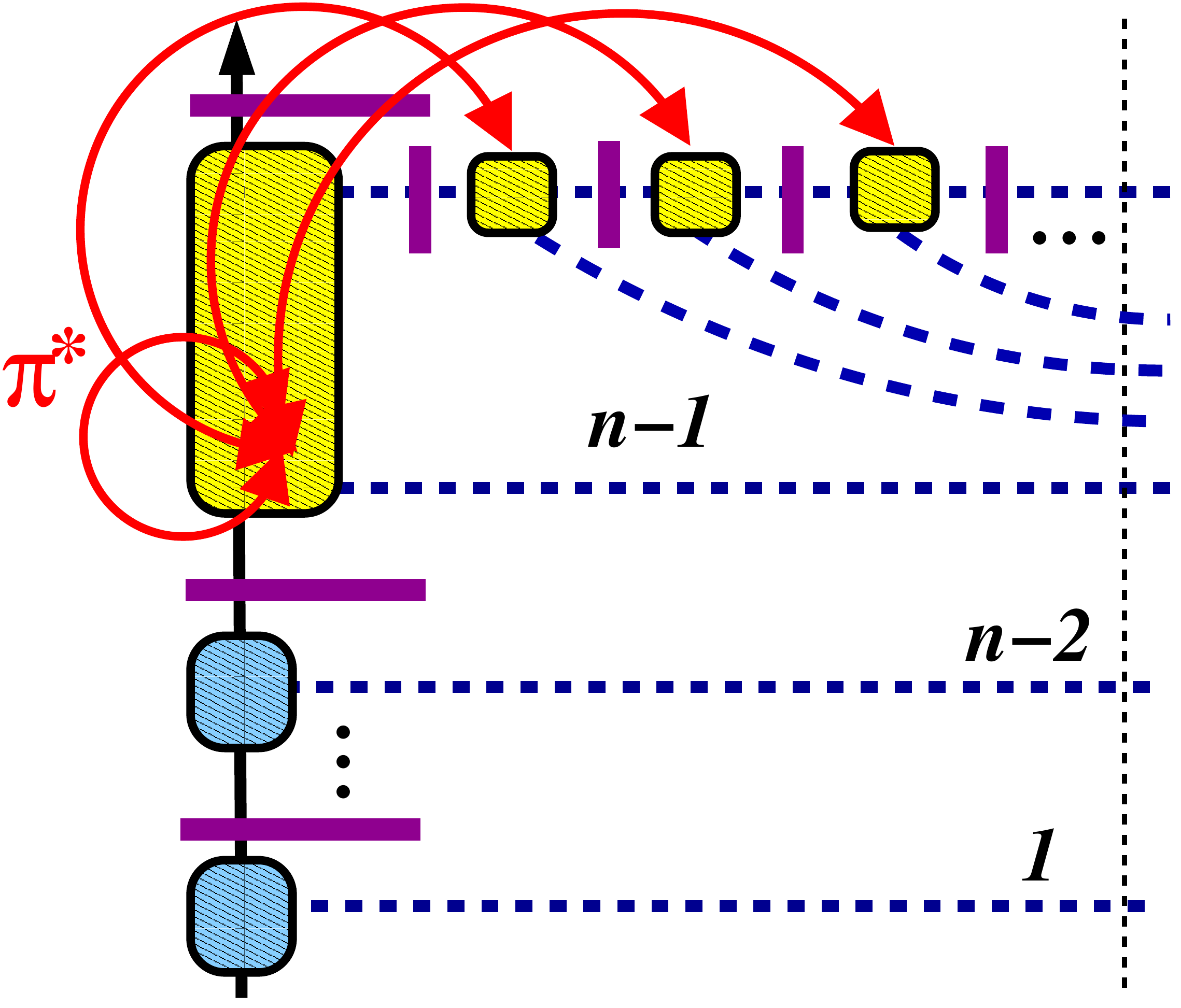}}
\end{equation}
where
$K^f=$\raisebox{-10pt}{\includegraphics[height=10mm]{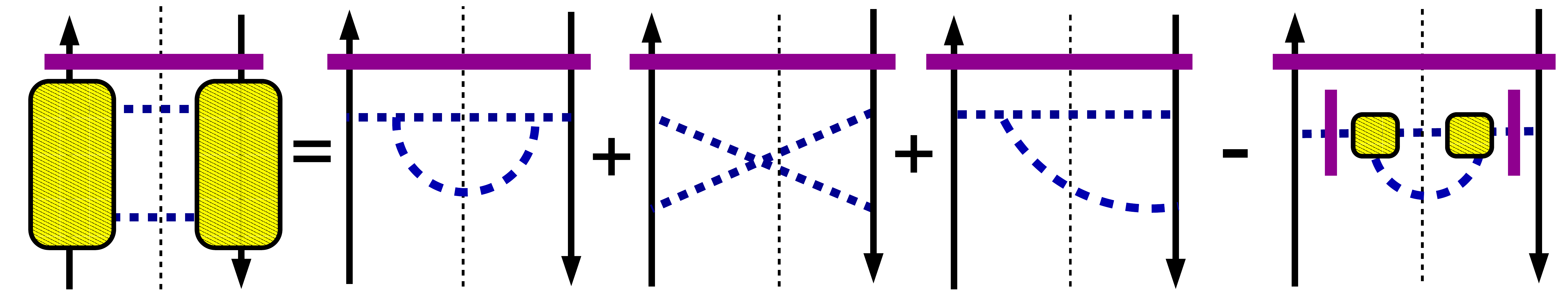}}
is the $\sim C_FC_A$ component in the above NLO MC weight.
Th leading FSR Sudakov double log part is subtracted in $K^f$.
The above sum in the MC weight was already tested for single
NLO insertion, giving rise to a well behaved weight distribution.
More testing is under way.

Summarizing, we report here on a small but essential part of the larger
project, with the aim of implementing NLO DGLAP evolution of the
parton distributions in fully unintegrated inclusive form
in the Monte Carlo with positive weights (weights equal one).
The main result of the ongoing study, so far limited to nonsinglet ePDF,
is that this is feasible.
Once completed, this project will lead to new type of the parton shower MC for
the initial state in LHC and other colliders with hadron beams.

\vspace{5mm}
\noindent
{\bf\large Acknowledgments}\\
Two of the authors (S.J and M.S.)
wish to thank Theory Unit of Physics Division, CERN,
for generous support during preparation of this work.


\end{document}